\documentclass[
superscriptaddress,
showpacs,preprintnumbers,nofootinbib,amsmath,amssymb,aps,prd,11pt, groupedaddress]{revtex4-2}
\usepackage{braket}
\usepackage{bm}
\usepackage{dcolumn}
\usepackage{cancel}
\usepackage[dvipdfmx,final]{graphicx}
\input{colordvi.tex}
\usepackage{longtable}
\usepackage{tabularx}
\usepackage{cleveref}
\usepackage{mhchem}
\usepackage{mathtools}
\usepackage[utf8]{inputenc}
\usepackage{color}
\usepackage{amsmath}
\usepackage{graphics}
\usepackage{epstopdf}
\usepackage{graphicx}
\usepackage{subfigure}
\usepackage{amsmath} 
\usepackage{amssymb} 
\usepackage{soul}

\makeatletter
\renewcommand{\fnum@table}{\textbf{\tablename~\thetable}}
\renewcommand{\fnum@figure}{\textbf{\figurename~\thefigure}}
\makeatother

\newcommand {\be}{\begin{equation}}
\newcommand {\ee}{\end{equation}}
\newcommand {\ba}{\begin{eqnarray}}
\newcommand {\ea}{\end{eqnarray}}

\begin{document}

\title{Exploring Solar Neutrino Oscillation Parameters with LSC at Yemilab and JUNO}

\author{Pouya Bakhti$^1$}
\email{pouya\_bakhti@jbnu.ac.kr}
\author{Meshkat Rajaee$^1$}
\email{meshkat@jbnu.ac.kr}
\author{Seon-Hee Seo$^2$}
\email{sunny.seo@ibs.re.kr}
\author{Seodong Shin$^{1,3}$}
\email{sshin@jbnu.ac.kr}
\affiliation{$^1$Laboratory for Symmetry and Structure of the Universe, Department of Physics, Jeonbuk National University, Jeonju, Jeonbuk 54896, Korea \\
$^2$Center for Underground Physics, Institute for Basic Science, 55 Expo-ro Yuseong-gu, Daejeon 34126, Korea \\
$^3$Particle Theory and Cosmology Group, Center for Theoretical Physics of the Universe, Institute for Basic Science (IBS), Daejeon, 34126, Korea
}

\begin{abstract}
We investigate the sensitivities of the liquid scintillator counter (LSC) at Yemilab and JUNO to solar neutrino oscillation parameters, focusing on $\theta_{12}$ and $\Delta m^2_{21}$.  
We compare the potential of JUNO with LSC at Yemilab utilizing both reactor and solar data
in determining those parameters.
We find that the solar neutrino data of LSC at Yemilab is highly sensitive to $\theta_{12}$ enabling its determination with exceptional precision.
Our study also reveals that if $\Delta m^2_{21}$ is larger, with a value close to the best fit value of KamLAND, JUNO reactor data will have about two times better precision than the reactor  LSC at Yemilab. 
On the other hand, if $\Delta m^2_{21}$ is smaller and closer to the best fit value of solar neutrino experiments, the precision of the reactor  LSC at Yemilab  will be better than JUNO. 
\end{abstract}

\maketitle

\section{Introduction}
\label{sec:intro}

Neutrinos are among the most elusive particles in the universe and hence studying their properties remains a key challenge in 
particle physics. One of the most intriguing phenomena in the study of neutrinos is neutrino oscillation, which refers to how neutrinos can transform between different flavors as they travel through space.
The study of neutrino oscillations has led to important insights into the properties of neutrinos and the fundamental laws of the universe. In particular, solar neutrino oscillations, which occur as neutrinos produced in the Sun travel to the Earth, have been the subject of intense research over the past few decades.

The solar neutrino problem 
arose from the observed deficit in 
the solar $\nu_e$
flux detected in various experiments such as Homestake \cite{Cleveland:1998nv}, Super-Kamiokande \cite{Super-Kamiokande:2005wtt, Super-Kamiokande:2008ecj, Super-Kamiokande:2010tar}, and Gallium experiments \cite{Kaether:2010ag, SAGE:2009eeu}.
In 2002, the SNO experiment \cite{SNO:2011hxd} resolved this issue by confirming the accuracy of the Standard Solar Models 
using the neutral current signal, which measured all active neutrino flavors. By comparing these results with the $\nu_e$ flux obtained from the charged current signal, SNO provided compelling evidence of electron neutrino conversion into other flavors.
Combining the SNO results with the data from the earlier experiments, along with the KamLAND reactor experiment \cite{KamLAND:2013rgu}, led to the discovery of the Large Mixing Angle (LMA) Mikheyev-Smirnov-Wolfenstein  (MSW) solution as the correct set of neutrino mixing parameters \cite{Wolfenstein:1977ue, Mikheyev:1985zog, Mikheev:1986wj}. 
 
In recent years, the Borexino experiment using liquid scintillator detectors has achieved impressive low energy thresholds in the sub-MeV region. As a result, precise measurements of solar $^7$Be  neutrinos have been obtained, along with the observation of solar $pep$ and $pp$ neutrinos for the first time \cite{BOREXINO:2020aww}. The observed results are consistent with the theoretical predictions of the MSW matter effect \cite{Wolfenstein:1977ue, Mikheyev:1985zog, Mikheev:1986wj}, providing further validation of this phenomenon in the context of neutrino oscillations. Indeed, there are still several key aspects of solar neutrinos that require further clarification. These include the need for improved determination of neutrino oscillation parameters, addressing the solar metallicity problem, and detailed analysis of the energy dependence in the low-energy region. 

In this paper, we focus on the sensitivities of the liquid scintillator counter (LSC) at Yemilab to solar neutrino oscillation parameters. 
Yemilab is  
a recently built underground laboratory located at Mt. Yemi in Jeongsun, Gangwon province, Korea.
A few kton scale liquid scintillator neutrino detector, LSC, is considered to be installed in Yemilab in the near future, which would play an important role in the study of low-energy neutrinos.  
In this study, we analyze the detector's ability to detect solar neutrinos as well as reactor neutrinos from the Hanul Nuclear Power Plant located at Uljin, which is 65 km away from Yemilab.

Our main goal is to investigate the potential of LSC at Yemilab in precisely determining the solar neutrino oscillation parameters, namely $\theta_{12}$ and $\Delta m^2_{21}$. We analyze the detector's ability to detect solar neutrinos and reactor neutrinos, and compare our results with the potential of the Jiangmen Underground Neutrino Observatory (JUNO)~\cite{JUNO:2015zny}, another leading neutrino detector in China, to determine these parameters. While the upcoming reactor experiment JUNO is expected to significantly enhance the measurement of certain oscillation parameters, such as solar parameters $ \theta_{12}$ and $\Delta m^2_{21}$, it is crucial for solar neutrino experiments to independently verify these measurements.  Precise determination of solar parameter $\Delta m^2_{21}$ can also play role in the determination of $\delta_{CP}$ in long-baseline experiments \cite{Denton:2023zwa}. The forthcoming solar neutrino experiment LSC at Yemilab, which benefits from a lower energy threshold compared to JUNO and the ability to detect $pp$ neutrinos with a larger fiducial volume compared to Borexino will provide higher statistics for determining the solar neutrino parameters. In this paper, we will elaborate on how LSC at Yemilab will be able to 
determine $\theta_{12}$ more precisely than JUNO. Furthermore, the combination of data from JUNO and LSC at Yemilab can offer an intriguing probe for the precise measurement of the solar parameters
and hence possible investigation of physics beyond the standard 3-neutrino oscillation \cite{Bakhti:2020hbz}.

Our results reveal that LSC at Yemilab exhibits a unique sensitivity to $\theta_{12}$, enabling precise determination of this parameter through the detection of solar neutrinos. Meanwhile, the value of $\Delta m^2_{21}$ can be determined with higher accuracy through the detection of reactor neutrinos. As we will elaborate more, if the value of $\Delta m^2_{21}$ is close to the best fit value of KamLAND, JUNO will have the best sensitivity to determine it, while if $\Delta m^2_{21}$ is closer to the best fit value of solar neutrino experiments, LSC at Yemilab can compete JUNO in measuring this value and provide a better measurement.
Therefore, combining data from both experiments
will yield the most accurate measurement of the solar neutrino oscillation parameters.

This paper is organized as follows.
We revisit the physics of solar neutrino propagation from the Sun to the Earth in Sec.~\ref{sec:sun} and neutrino oscillation in reactors in Sec.~\ref{sec:reactor}.
The detailed information of LSC at Yemilab and JUNO is summarized in Sec.~\ref{sec:exp}.
Our main analysis results with proper plots are explained in Sec.~\ref{sec:results}.
Finally, we summarize our conclusions and discuss future prospects in Sec.~\ref{sec:conclusions}.

\section{Solar Neutrino Propagation from the Sun to the Earth}
\label{sec:sun}

Neutrino oscillation in matter is a phenomenon that occurs when neutrinos pass through a medium with a constant or varying density, such as the Sun or the Earth. The flavor of the neutrinos can change due to their interactions with the surrounding matter while propagating. This is known as MSW effect~\cite{Wolfenstein:1977ue, Mikheyev:1985zog, Mikheev:1986wj}. The MSW effect can have a significant impact on the observed solar neutrino flux depending on the density profile of the Sun. In the case of solar neutrinos, the flavor conversion is most sensitive to the mixing angle $\theta_{12}$ and the mass-squared difference $\Delta m^2_{21}$, which governs the oscillation between electron neutrinos and other neutrino flavors.

The evolution of a neutrino flavor state is described by the Shr\"odinger-like equation:
\begin{align}
i\frac{d|\nu_f\rangle}{dx} = H|\nu_f\rangle = (H_0 + V)|\nu_f\rangle\,,
\label{eq:Hamiltonian}
\end{align}
where $H$ is the total Hamiltonian, $H_0$ is the vacuum Hamiltonian, and $V = \mathrm{diag}(V_e,0,0)$ is the diagonal matrix of matter potentials with $V_e = \sqrt{2}G_Fn_e$. Here, $G_F$ is the Fermi constant and $n_e$ is the number density of electrons.
The parameter $x$ denotes the oscillation length approximately the same as the propagation time. Note that the Hamiltonian is not diagonal in the flavor basis.

The evolution of flavor neutrino states is described in terms of the instantaneous eigenstates of the Hamiltonian in matter, denoted as $|\nu_m\rangle \equiv (\nu_{1m},\nu_{2m},\nu_{3m})^T$. The relationship between these eigenstates and the flavor states is expressed through the mixing matrix in matter, denoted as $U_m$:
\begin{align}
|\nu_f\rangle = U_m|\nu_m\rangle\,.
\end{align}
The matrix $U_m$ diagonalizes the Hamiltonian:
\begin{align}
U_m^\dagger H U_m =  \mathrm{diag}(H_{1m},H_{2m},H_{3m})\,,
\end{align}
where $H_{im}$ are the eigenvalues of the Hamiltonian.

The propagation and flavor evolution of solar neutrinos can be studied in three distinct stages: 1) propagation inside the Sun, 2) propagation from the Sun to the Earth, and 3) propagation inside the Earth. When an electron neutrino is produced in the core of the Sun, it propagates through 
space inside the Sun as a combination of different neutrino mass states, known as eigenstates of the Hamiltonian
in Eq.~(\ref{eq:Hamiltonian})

While propagating through the space between the Sun and the Earth, the high energy neutrinos lose coherence and no oscillations occur.~\footnote{ The low energy neutrinos have already lost the coherence inside the Sun.} 
However, as they enter the Earth, the mass states split into the eigenstates of the Hamiltonian in the Earth's matter and begin
to oscillate as they travel through the Earth towards a detector. In the following, we will elaborate on each phase independently.

As solar neutrinos travel from their production point in the Sun to the Earth, 
the wave packets associated with different 
mass
eigenstates will separate due to their distinct group velocities, leading to the loss of propagation coherence. Moreover, each wave packet spreads as a consequence of having different momenta. However, it is important to note that although the spread is present, it remains smaller compared to the separation between the eigenstates \cite{Kersten:2015kio}. Thus, this spreading does not significantly impact the coherence condition.
As a result, the fluxes of the mass eigenstates of solar neutrinos arrive at the Earth incoherently. The probability of detecting a $\nu_e$ at arrival time $t$ 
on the {\it surface} of the Earth 
is given by \cite{Maltoni:2015kca}
\begin{align}
\label{eq:patearth}
P_{ee}^{\rm sur} = |\langle \nu_e | \nu (t) \rangle |^2
= \sum_j |U^m_{ej}(n_e^0)|^2 |U_{ej}|^2 \,,
\end{align}
where $U^m_{ej}(n_e^0)$ are the elements of mixing matrix in the
production point with density $n_e^0$.
Using the standard parametrization of the mixing matrix, $P_{ee}^{\rm sur}$ can be written as \cite{Maltoni:2015kca, Bakhti:2020tcj}
\begin{align} 
  \label{eq:nueday1}
  P_{ee}^{\rm sur} = \frac{1}{2} c_{13}^2 c_{13}^{m2} (1 + \cos 2\theta_{12} \cos 2 \theta_{12}^m)
  + s_{13}^2 s_{13}^{m2} \,.
\end{align} 
The 1-2 mixing angle $\theta_{12}^m$ is determined by
\begin{align}
\label{costheta}
  \cos 2 \theta_{12}^m =
  \frac{\cos 2\theta_{12} - c_{13}^2 \epsilon_{12}}{\sqrt{(\cos 2\theta_{12}
      - c_{13}^2 \epsilon_{12})^2 + \sin^2 2\theta_{12}}} \,,
\end{align} 
where
\begin{align}
  \label{eq:eps12}
  \epsilon_{12} \equiv \frac{A_{CC}}{\Delta m^2 _{21}} \,,
\end{align}
with $A_{CC}=2V_e E$. Notice that matter effect is important for high energies where $\epsilon_{12}$ is not negligible.

Finally, at the time solar neutrinos reach the Earth's surface, they undergo a flavor conversion due to the presence of matter. 
The probability of detecting an electron neutrino now becomes:

\begin{align}
P_{ee}=\sum_{j} |U_{ej}^m (n^0 _e )|^2 \cdot P_{je} \,,
\end{align}
where $P_{je}$ are the probabilities of {\it oscillatory} 
transitions $\nu_j \rightarrow \nu_e$ in the Earth matter \cite{Ioannisian:2017dkx,  Bakhti:2020tcj}, 
while 
$|U_{ej}|^2$ 
 in Eq.~(\ref{eq:patearth}) is the square of the {\it fixed} component of the mixing matrix.

\section{Neutrino Oscillation in Reactors}
\label{sec:reactor}

The mechanism of neutrino oscillation can be explained by the fact that the three neutrino flavors are not distinct states of definite mass, but are quantum mechanical superpositions of three different mass states. As a result, as neutrinos travel through 
(some) space, they can undergo a transformation among those different mass eigenstates, leading to oscillations into different flavor eigenstates at the time of detection.
The PMNS matrix, also known as the Pontecorvo-Maki-Nakagawa-Sakata matrix, is a unitary matrix that describes the mixing of neutrino flavor states (electron, muon, and tau) with the neutrino mass eigenstates ($\nu_1$, $\nu_2$, and $\nu_3$). The PMNS matrix can be parameterized with three mixing angles, $\theta_{12}$, $\theta_{13}$ and $\theta_{23}$ and one $\delta_{CP}$ phase if the neutrinos are Dirac fermions. 

The probability of a produced $\alpha$ flavor neutrino 
$\nu_\alpha$ being 
detected as a $\beta$ flavor neutrino  $\nu_\beta$ is given by:
\begin{align}
\label{neutrino_oscillation}
P_{\nu_{\alpha} \rightarrow \nu_{\beta}}(t) = \sum_{k,j} U_{\alpha k}^* U_{\beta k} U_{\alpha j} U_{\beta j}^* \exp\left( -\frac{i \Delta m_{kj}^2 L}{2E} \right),
\end{align}
where $E$ is the energy of the neutrino, $U_{\alpha k}$ are the elements of the PMNS matrix and  $\Delta m^2_{ij}$ is the squared-mass differences between mass eigenstate of $i$ and $j$, and $L$ is the baseline. 
As stated in the previous section, the presence of matter can significantly affect the neutrino oscillations, particularly for neutrinos with high energies and high matter density where $\epsilon_{12} \equiv  A_{CC}  / \Delta m^2_{21}$ is significant.  The matter effect is not important for reactor neutrinos with energies of $\mathcal O ({\rm MeV})$ which propagate through the Earth crust with density of 2.6~$gr/cm^3$, due to the fact that $\epsilon_{12}$ is negligible.

In reactor neutrino experiments, electron antineutrinos ($\bar{\nu}_e$) are produced and detected. The oscillation probability $P(\bar{\nu}_e \to \bar{\nu}_e)$ is expressed as follows:
 \be \label{Pee} P(\bar{\nu}_e \to
\bar{\nu}_e) =\left| |U_{e1}|^2 +|U_{e2}|^2e^{i
\Delta_{21}}+|U_{e3}|^2e^{i \Delta_{31}}\right|^2= \left|
c_{12}^2c_{13}^2+ s_{12}^2 c_{13}^2 e^{i\Delta_{21}}+s_{13}^2 e^{i
\Delta_{31}}\right|^2 \ee
$$ = c_{13}^4 (1-\sin^2 2\theta_{12}
\sin^2\frac{\Delta_{21}}{2})+s_{13}^4+2s_{13}^2c_{13}^2[\cos
\Delta_{31}(c_{12}^2+s_{12}^2 \cos \Delta_{21})+s_{12}^2\sin
\Delta_{31}\sin \Delta_{21}]\,,$$
where $\Delta_{ij}=\Delta m_{ij}^2L/(2E_\nu)$. 
For reactor experiments with $\mathcal{O}(1~{\rm km})$ baseline, such as Daya Bay~\cite{DayaBay:2012fng}, RENO~\cite{RENO:2012mkc}, or Double-CHOOZ~\cite{DoubleChooz:2019qbj}, the value of $\Delta m^2 _{21}$ can be effectively approximated as zero due to the relatively small baseline length. Consequently, this approximation results in reduced sensitivity when measuring the mixing angle $\theta_{12}$. However, it's important to note that these experiments retain their sensitivity to larger values of $\Delta m^2 _{21}$~\cite{Seo:2018rrb, Hernandez-Cabezudo:2019qko}. 
For neutrino oscillation with one km baseline reactor neutrinos, the key parameters are $\theta_{13}$ and $\Delta m^2_{31}$. 
On the other hand, the value $\Delta_{21}$ can be non-negligible in longer baseline experiments such as LSC at Yemilab (from Hanul power plant), JUNO, and KamLAND. 
Reactor neutrinos offer significant advantages for the study of neutrino oscillation due to their large flux and well-characterized energy spectrum. Moreover, the relatively short baselines between reactor neutrino sources and detectors, typically spanning tens of kilometers, make them exceptionally well-suited for precise measurements of oscillation parameters.
Reactor neutrino experiments with shorter baselines, such as Daya Bay and RENO have successfully delivered precise measurements of the mixing angle $\theta_{13}$~\cite{DayaBay:2012fng,RENO:2012mkc}. 
On the other hand, experiments with longer baselines hold the potential to unravel crucial information regarding other significant oscillation parameters, including $\theta_{12}$, $\Delta m^2_{21}$, and $\Delta m^2_{31}$. 
In the following section, we will provide detailed explanations of the two reactor experiments utilized in this study: LSC at Yemilab and JUNO.

\section{Details of LSC at Yemilab and JUNO experiments}
\label{sec:exp}

Yemilab is situated 1 km deep in Handuk iron mine, Jeongsun-gun, Gangwon-do, Korea. 
The primary research programs to start in 2023 - 2024 at Yemilab are the dark matter direct detection with COSINE-200 and neutrinoless double-beta decay search with AMoRE-II. LSC is a multi-purpose detector covering from astroparticle to particle physics, and if funded, its main physics would be to precisely measure solar neutrinos, search for sterile neutrinos using IsoDAR and/or radioactive source(s), and search for dark photons using a linear accelerator (LINAC)~\cite{HSLPPC}.

One of the key advantages of a liquid scintillator detector LSC is its remarkable ability to detect a wide range of neutrino energies, spanning from a few 100 keV to several GeV. This exceptional energy range makes it an ideal detector for studying solar neutrinos, which exhibit energies ranging from a few 100~keV to less than twenty MeV. The detector  consists of a 2.26 kton liquid scintillator housed within an acrylic 
cylinder
vessel with dimensions of 15~m in diameter and 15~m in height. The buffer region surrounding the scintillator is filled with mineral oil, weighing 1.14 ktons, contained within a stainless steel vessel measuring 17~m in diameter and 17~m in height. Additionally, the veto region is filled with purified water, amounting to 2.41 ktons, in a stainless steel vessel (or no vessel but a lining of the cavity) with dimensions of 20~m in diameter and 20~m in height.

Yemilab 
is situated approximately 65~km away from the Hanul Nuclear Power Plant at Uljin, Gyeongsangbuk-do, Korea. This enables Yemilab to utilize reactor neutrinos emitted by the power plant as a valuable calibration source for the 
neutrino detectors.
By detecting and analyzing both solar and reactor neutrinos, 
LSC at Yemilab has an excellent capability of investigating neutrino oscillations to accurately determine the oscillation parameters of solar neutrinos, namely $\theta_{12}$ and $\Delta m^2_{21}$. 
An article to describe more details on LSC, its design, and physics potential is being prepared separately~\cite{LSCwhite}.

In our analysis of reactor neutrino detection, we have considered a fiducial volume of 2 kton for LSC. The reactor complex at 
Hanul nuclear power plant
consists of 
eight
reactors, namely Hanul-1 to Hanul-6, and Shin-Hanul-1 and Shin-Hanul-2
(from this September), with a total thermal power output of 24.816 GW$_{th}$. Among these reactors, Hanul-1 and Hanul-2 have a power output of 2.775 GW$_{th}$ each, while Hanul-3 to Hanul-6 have a power output of 2.825 GW$_{th}$ each. In addition to these reactors, the complex includes two new reactors, Shin-Hanul-1 and Shin-Hanul-2, with a power output of 3.983 GW$_{th}$ each ~\cite{LSCwhite}.
The distance between LSC at Yemilab and the reactors is approximately 65 km. 
In terms of the background for LSC reactor neutrino detection, we have taken into account a level of background similar to that of JUNO, but with 10 times smaller volume of LSC and 2 times better shielding of muons due to more overburden for LSC. This results in 20 times fewer background events in LSC compared to JUNO.

We have assumed a detector efficiency of $90\%$, normalization uncertainty of $0.8\%$, energy calibration error of $0.5\%$, and have adopted other experimental details consistent with the reference \cite{Huber:2003pm}. To carry out our analysis we have used GLoBES software \cite{Huber:2004ka, Huber:2007ji}. For the statistical inferences, we have considered the Asimov data set approximation.

In our analysis of solar neutrinos at LSC, the energy threshold is taken as 0.2~MeV. This is because the cosmogenic $^{11}$C background events below 0.2~MeV are huge, several orders of magnitude larger than $pp$ neutrino signal.
   Above
0.2 MeV,  dominant backgrounds are cosmogenic muon-induced backgrounds and  internal and environmental radioactive backgrounds. 
To account for these backgrounds, we have adopted the same background modeling approach as the Jinping
experiment \cite{Jinping:2016iiq}.
In our analysis of solar neutrinos, we have conservatively assumed a fiducial volume of 1 kton for LSC at Yemilab.

JUNO is an advanced liquid scintillator detector situated in southern China, at a distance of approximately 53 km from the Yangjiang and Taishan nuclear power plants.  
The JUNO detector is composed of a central acrylic sphere measuring 35~m in diameter, which is filled with 20 kton of liquid scintillator. The scintillator is doped with gadolinium to enhance the detection of neutron interactions. Furthermore, 20,000 (25,600) 20-inch (3-inch) photomultiplier tubes (PMTs) are mounted on a stainless steel lattice structure outside the acrylic sphere vessel, enabling the detection of light emitted by neutrino interactions within the scintillator~\cite{JUNO:2015zny}.

One of the primary goals of JUNO is to investigate neutrino oscillations utilizing the reactor neutrinos generated by nearby nuclear power plants. Reactor neutrinos possess a well-characterized energy spectrum and flavor composition, rendering them an exceptional resource for studying neutrino oscillations. The key focus of JUNO is to determine the mass ordering of neutrinos, a crucial aspect in understanding their fundamental properties. Additionally, JUNO offers the advantage of measuring the oscillation parameters $\theta_{12}$ and $\Delta m^2_{21}$ with remarkable precision. With its substantial detector size and close proximity to the nuclear power plants, JUNO is anticipated to achieve a statistical uncertainty on $\sin^22\theta_{12}$ that is approximately four times superior to the current best measurement achieved by the KamLAND experiment~\cite{JUNO:2015zny}.

In addition to reactor neutrinos, JUNO is also sensitive to solar, atmospheric, and supernova neutrinos. With its exceptional detection efficiency, 
outstanding energy resolution (ranging from 3$\%$ to $3.5\%$), and minimal background levels, the JUNO detector serves as a powerful instrument for scrutinizing neutrinos from various sources. 
While JUNO may not detect the lowest energy $pp$ neutrinos, it possesses the capability to measure the higher energy $^8$B neutrinos with a threshold of 3.5 MeV. Moreover, JUNO is actively exploring methods to mitigate the background originating from $^{210}$Bi in order to detect $^7$Be neutrinos, which encompass energies of 0.384 MeV and 0.862 MeV \cite{JUNO:2015zny}.

The study of solar neutrinos stands as a crucial component of JUNO's scientific agenda. By precisely measuring the flux and energy spectrum of solar neutrinos, JUNO can furnish invaluable insights into the fusion processes that energize the Sun, as well as neutrino oscillations transpiring within the high-density environment of the solar interior. 
In particular, JUNO holds the potential to make substantial advancements in our comprehension of the solar abundance problem and other discrepancies, which pertains to the persistent discrepancy between the anticipated and observed flux of solar neutrinos. Through the precise detection of $^8$B neutrinos and the measurement of other solar neutrino fluxes, JUNO may provide better constraints on solar models and neutrino oscillation parameters.
JUNO is also expected to detect a small number of $hep$ neutrinos, which are produced by a rare fusion process in the Sun. The detection of $hep$ neutrinos, which have a maximum energy of approximately 18.8 MeV, will provide a unique probe of the high-energy neutrino physics of the solar core.

For the analysis of JUNO reactor neutrinos, we have adopted the same experimental details as described in references \cite{Bakhti:2014pva, Bakhti:2013ora}. We have assumed a selection efficiency of $73\%$ for electron anti-neutrino \cite{JUNO:2015zny}.
The primary sources of background in our analysis include cosmogenic background~\cite{Grassi:2014hxa}, accidental background, geo-neutrino background, and the contribution from $^{13}{\rm C}(\alpha,n)^{16}{\rm O}$. We have considered a flux normalization uncertainty of $5\%$. The scintillator detector employed in JUNO will have a fiducial mass of 20 kton and an energy resolution of $3\%$ is assumed in our analysis \cite{JUNO:2015zny}.

In the analysis of JUNO low-energy solar neutrinos, we have considered the detection of $^7$Be neutrinos and a very small contribution from $pp$ neutrinos with electron kinetic energy greater than 0.2 MeV. Below 0.2 MeV, the presence of a significant $^{14}$C background poses a challenge as it is several orders of magnitude higher than the solar neutrino signals. The number of events from other low-energy solar neutrinos is much smaller compared to the background. The main sources of background at low energies include $^{210}$Bi, $^{85}$Kr, $^{40}$K, $^{14}$C, $^{238}$U, $^{232}$Th, and $^{11}$C. In our analysis, we have considered two scenarios for the background: an ordinary case where the background is reduced compared to KamLAND's solar neutrino detection, and an ideal radiopurity assumption that is comparable to the lower background observed in Borexino phase I.
For the detection of high-energy $^8$B neutrinos, the main sources of background are cosmogenic isotopes such as $^{11}$C, $^{10}$C, and $^{11}$Be. Below 3.5 MeV, both $^{11}$C and $^{10}$C contribute a large number of background events. Above 3.5 MeV, the only significant source of background is $^{11}$Be, which may be half of the number of events or comparable to the $^8$B solar neutrino signal \cite{JUNO:2015zny, JUNO:2020hqc}.

Additionally, neutron captures on steel produce gamma rays with energies of 6 MeV and 8.5 MeV, which can penetrate the center of the JUNO detector~\cite{JUNO:2015zny}. To mitigate this background, reducing the fiducial volume to less than half of the detector's size is necessary. In our analysis, we have considered a 20 kton detector for the detection of low-energy solar neutrinos ($^7$Be neutrinos), and a   16 kton detector for the detection of $^8$B neutrinos. 
 We have assumed efficiencies of $100\%$ and $50\%$ for $^7$Be and $^8$B neutrinos, respectively~\cite{JUNO:2020hqc}. As we will elaborate on, our calculations demonstrate that for the detection of $hep$ neutrinos with electron kinetic energies greater than 14.8 MeV, using a 20 kton fiducial volume, we anticipate six events after ten years of data collection while we expect the detection of only one $^8$B solar neutrino in this energy range.   However, the actual fiducial volume could be smaller than this.
The details of the fiducial volume we considered for the analysis and the main backgrounds for each solar and reactor neutrino are tabulated in Table~\ref{table:fiducial}.

\begin{table}[htbp]
\centering
\begin{tabular}{|l|l|p{5.5cm}|}
\hline
\textbf{Experiment} & ~~\textbf{Fiducial Volume considered} ~~ &  \textbf{Background} ~~ \\
\hline
JUNO (reactor) & 20 kton & Cosmogenic background, accidental background, Geoneutrino background, $^{13}$C$(\alpha,n)^{16}$O \\
\hline
JUNO (Solar $^7$Be) & 20 kton & Cosmogenic backgrounds, intrinsic radioactive background \\
\hline
JUNO (Solar $^8$B) & 16 kton & Cosmogenic backgrounds, intrinsic radioactive background, and neutron capture

\\
\hline
LSC at Yemilab (reactor) & 2 kton & Cosmogenic background, accidental background, Geoneutrino background, $^{13}$C$(\alpha,n)^{16}$O\\
\hline
LSC at Yemilab (solar) & 1 kton &  Cosmogenic backgrounds, intrinsic radioactive background\\
\hline
\end{tabular}
\caption{
The fiducial volume we considered for the analysis and the main backgrounds for each solar and reactor neutrino in LSC at Yemilab and JUNO.
}
\label{table:fiducial}
\end{table}

\section{Results}
\label{sec:results}

In this section, we present our 
analysis results of the sensitivities of LSC at Yemilab and JUNO on the solar neutrino oscillation parameters. 
First, we will discuss the oscillation probabilities for reactor neutrinos and solar neutrinos. Subsequently, we will provide the number of events for the reactor neutrino data and solar neutrino data. Finally, we will illustrate the sensitivities of the two experiments 
on solar neutrino oscillation parameters, specifically focusing on $\Delta m^2_{21}$ and $\theta_{12}$. In our analysis, we focus on the survival probabilities of electron (anti-)neutrinos due to the exclusive production of electron (anti-)neutrinos by both reactor neutrinos and solar neutrinos. The baselines for reactor neutrinos are set to be 52.5 km for JUNO and 65 km for LSC at Yemilab, respectively. To determine the oscillation parameters, we adopt the current best fit values obtained from the KamLAND data: $\Delta m^2_{21} = 7.54 \times 10^{-5}$ eV$^2$ and $\sin^2\theta_{12} = 0.316$. Additionally, we consider the best fit values from the combination of Super-Kamiokande (SK) and Sudbury Neutrino Observatory (SNO) solar data, which yield $\Delta m^2_{21} = 6.11 \times 10^{-5}$ eV$^2$ and $\sin^2\theta_{12} = 0.306$ \cite{nakajima2020recent}. 

\subsection{Oscillation probabilities and the expected number of events}
\label{sec:osc-results}

\begin{figure}[h]
\begin{center}
\includegraphics[width=0.49 \textwidth]{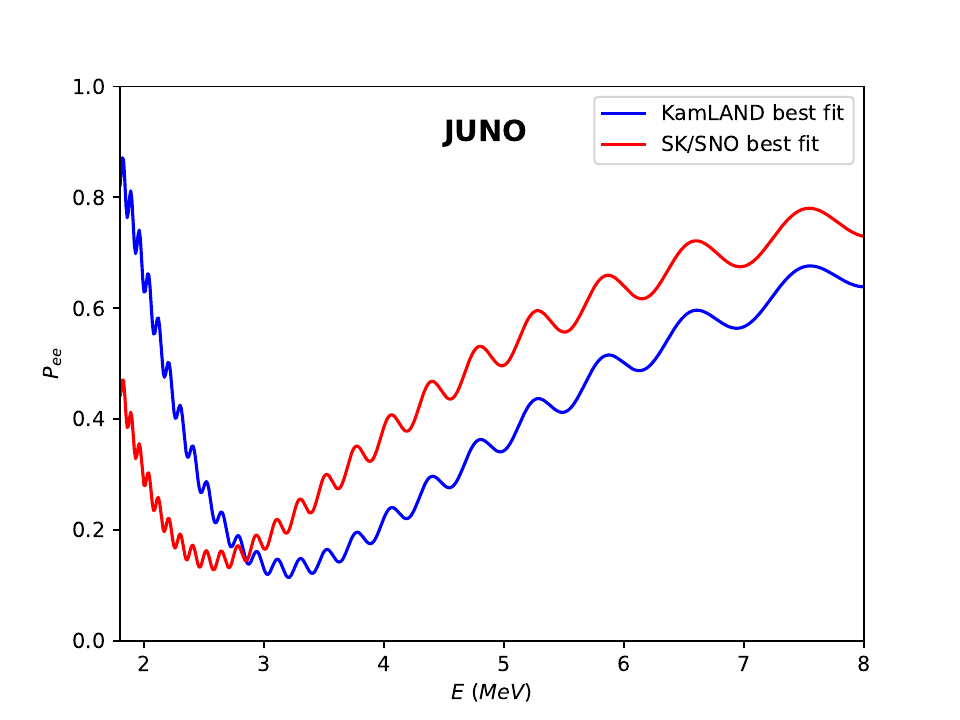}
\includegraphics[width=0.49 \textwidth]{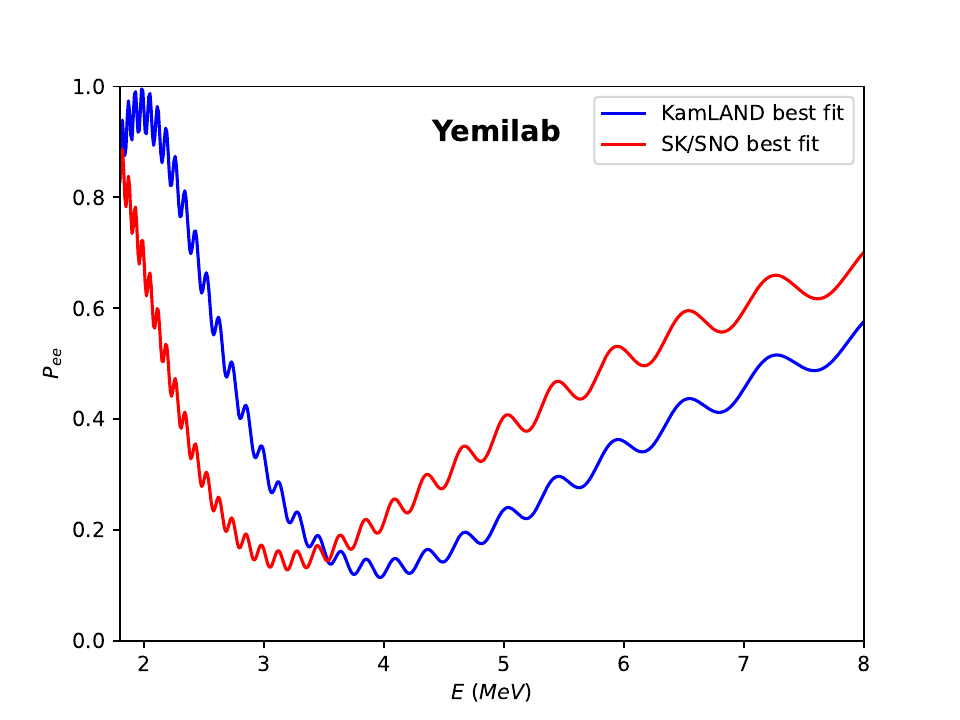}
\end{center}
\caption{\label{pee_reactor} Reactor neutrino survival probability, $P_{ee}$, as a function of neutrino energy for different values of solar parameters, $\Delta m^2_{21}$ and $\theta_{12}$. The blue curves correspond to the best fit values of KamLAND data, while the red curves correspond to the best fit values of SK/SNO solar neutrino data. These curves represent the oscillation probabilities for baselines of 52.5 km (JUNO baseline) and 65 km (Yemilab baseline).
}
\end{figure}

Figure~\ref{pee_reactor} shows the electron neutrino survival probabilities $P_{ee}$ as a function of energy for two different baselines. The blue curves correspond to the best fit values of KamLAND data ($\Delta m^2_{21} =7.54\times 10^{-5}$ eV$^2$ and $\sin^2\theta_{12}=0.316$), while the red curves correspond to the best fit values of SK/SNO data ($\Delta m^2_{21} =6.11\times 10^{-5}$ eV$^2$ and $\sin^2\theta_{12}=0.306$) \cite{nakajima2020recent}. As expected, the oscillation probabilities at LSC at Yemilab (65 km) and JUNO (52.5 km) exhibit distinct patterns due to the difference in baselines. The curves demonstrate that the electron neutrino survival probabilities at both baselines are influenced by the values of $\Delta m^2_{21}$ and $\theta_{12}$. In our analysis, we assume a normal mass ordering and fix $\theta_{13}$ at the best fit value of the Daya Bay experiment \cite{DayaBay:2018yms}   
and the value of $\Delta m^2_{31}$ to the best fit value of the nu-fit analysis \cite{Esteban:2020cvm}.

\begin{figure}[h]
\begin{center}
\includegraphics[width=0.7 \textwidth]{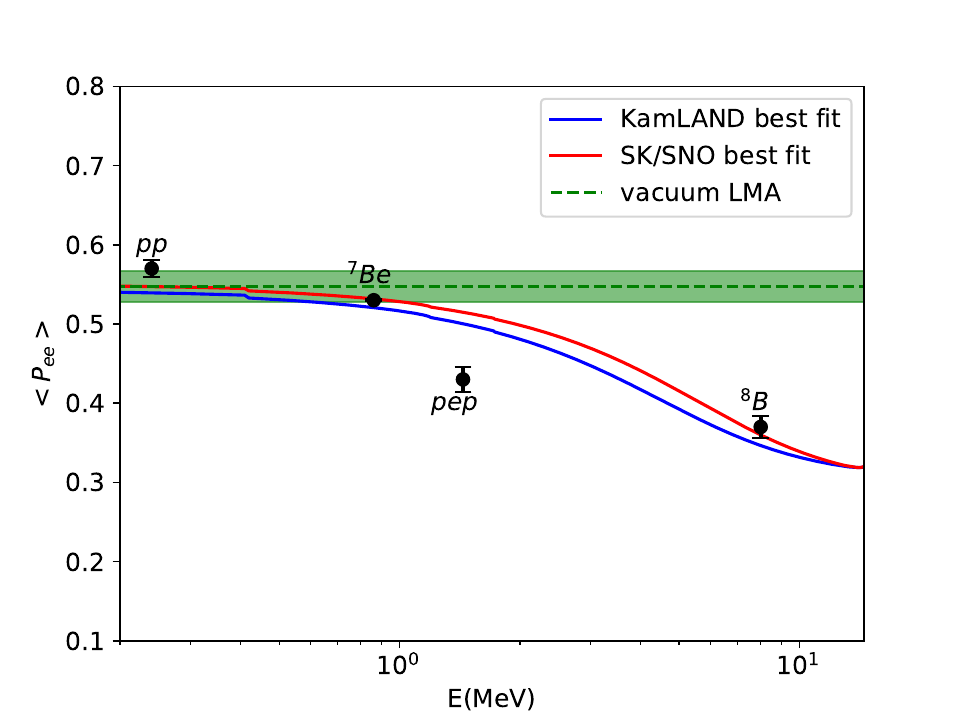}
\end{center}
\caption{\label{pee_solar} Solar neutrino survival probability, $P_{ee}$, as a function of neutrino energy for different values of $\Delta m^2_{21}$ and $\theta_{12}$, corresponding to the best fit values of KamLAND data (blue curve) and SK/SNO solar neutrino data (red curve).  Data points are expected from LSC at Yemilab for ten years of operation (statistical error only), using Borexino-measured values~\cite{BOREXINO:2020aww} as the central values.}
\end{figure}

Figure~\ref{pee_solar} depicts the electron neutrino survival probabilities $P_{ee}$ as a function of energy for solar neutrinos. We have taken into account various aspects related to solar neutrinos and the Sun, such as the density of the Sun and the distribution functions of neutrino production in the $pp$ chains and CNO cycle, as described in Ref.~\cite{Bahcall:2004pz}. The blue curves correspond to the best fit values of KamLAND data ($\Delta m^2_{21} =7.54\times 10^{-5}$ eV$^2$ and $\sin^2\theta_{12}=0.316$), while the red curves correspond to the best fit values of SK/SNO solar neutrino data ($\Delta m^2_{21} =6.11\times 10^{-5}$ eV$^2$ and $\sin^2\theta_{12}=0.306$)~\cite{nakajima2020recent}. 
Note that the quantity $P_{ee}$ represents the ratio of the averaged electron neutrino flux to the sum of all fluxes ($P_{ee}=\frac{\phi_{CC}}{\phi_{NC}}$). 
The data points are set to the best fit values of Borexino results for the $pp$, $^7$Be, $pep$, and $^8$B fluxes, assuming the GS98 solar model \cite{Vinyoles:2016djt}. 
The error bars indicate the statistical uncertainties in the ten years of running LSC at Yemilab.   Notice that if $\Delta m^2_{21}$ lies within the range $5 \times 10^{-5} < \Delta m^2_{21} < 9 \times 10^{-5}$, the oscillation of $pp$ neutrinos is primarily sensitive to the value of $\theta_{12}$. This sensitivity arises due to vacuum oscillations occurring in the energy range of approximately $0.2$ to $0.4$ MeV (see Eq.~(\ref{eq:nueday1})).  In the case of $^7$Be neutrinos, the matter effect becomes more significant due to their higher energy. As a result, they exhibit sensitivity to both $\Delta m^2_{21}$ and $\theta_{12}$. Also, as can be observed from Figure~\ref{pee_solar}, $^7$Be is close to the vacuum solution with only a small deviation, dependent on the $\Delta m^2_{21}$. 
Moreover, the resonance occurs specifically for $^8$B neutrinos, making them essential for precise measurements of both $\Delta m^2_{21}$ and $\theta_{12}$.

\begin{figure}[h]
\begin{center}
\includegraphics[width=0.49 \textwidth]{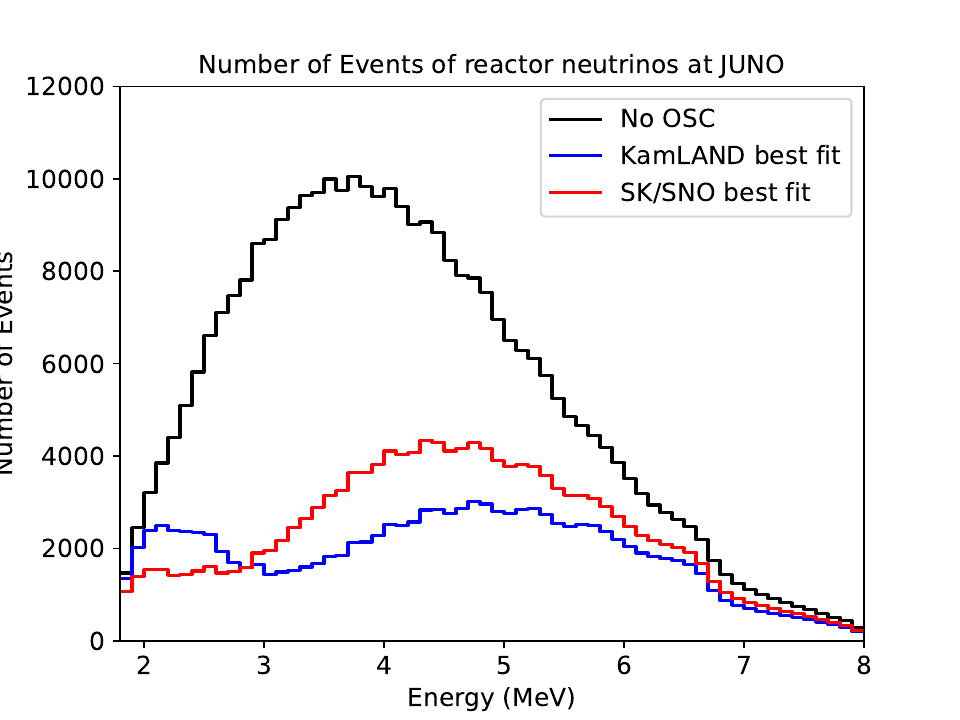}
\includegraphics[width=0.49 \textwidth]{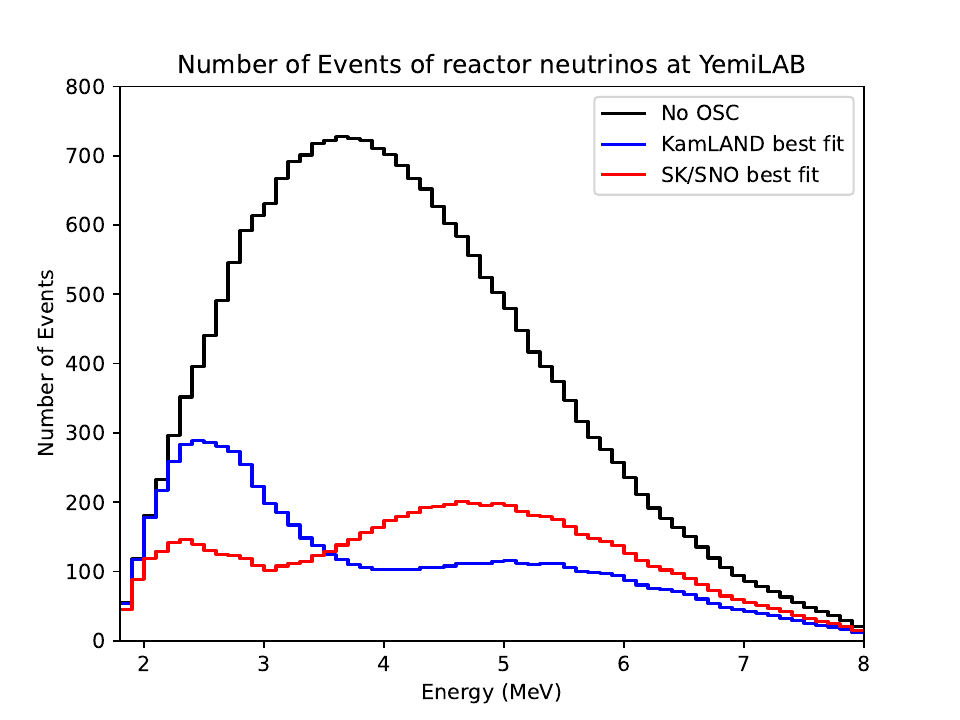}
\end{center}
\caption{ The number of events per bin (0.1 MeV) as a function of energy for two different sets of $\Delta m^2_{21}$ and $\theta_{12}$. The blue curves correspond to the best fit values of Kamland data, while the red curves correspond to the best fit values of SK/SNO solar neutrino data. The calculations are based on a data-taking period of ten years and consider two different baselines: 52.5 km, corresponding to the JUNO baseline, and 65 km, corresponding to the Yemilab baseline.}\label{noe_reactor}
\end{figure}

We present the number of events of reactor neutrinos detectable by JUNO and LSC at Yemilab in Fig.~\ref{noe_reactor}. The black curve represents the case of no-oscillation, while the blue and red curves correspond to the oscillation scenarios characterized by $\Delta m^2_{21}$ and $\theta_{12}$, based on the best fit values of KamLAND data and the best fit values of SK/SNO solar neutrino data, respectively. Our calculations assume ten years of data collection for both LSC at Yemilab and JUNO reactor neutrino experiments. The figure shows that the number of events of reactor neutrinos without oscillation is approximately twenty times higher at JUNO compared to LSC at Yemilab due to the differences in the experimental setup.
JUNO, with a reactor power of  26.6 GW$_{th}$, a 20 kton detector, and a 52.5 km baseline, outperforms LSC at Yemilab, which has a reactor power of 24.8 GW$_{th}$, a 2 kton detector, and a 65 km baseline
 \footnote{ The ratio of the number of events at JUNO to LSC at Yemilab without oscillation can be estimated as $(26.6/24.8)\times(20/2)\times(65/52.5)^2$.}.
We can see the oscillation behavior of the blue and red solid lines.  
Note that in the case of no oscillation (black curve), the maximum number of events occurs at the energies around $3-5$ MeV. In the right panel of Fig.~\ref{pee_reactor},  we can observe that LSC at Yemilab is more sensitive to $\Delta m^2 _{21}$ because the maximum flavor conversion also occurs around $3-4$ MeV, for both KamLAND and SK/SNO best fit values. On the other hand, for JUNO, the maximum occurs in the energy range of $2-3$ MeV, where the number of events is lower by about 30$\%$ compared to the SK/SNO best fit values. Consequently, JUNO demonstrates lower sensitivity compared to Yemilab.

\begin{figure}[h]
\begin{center}
\includegraphics[width=0.52 \textwidth]{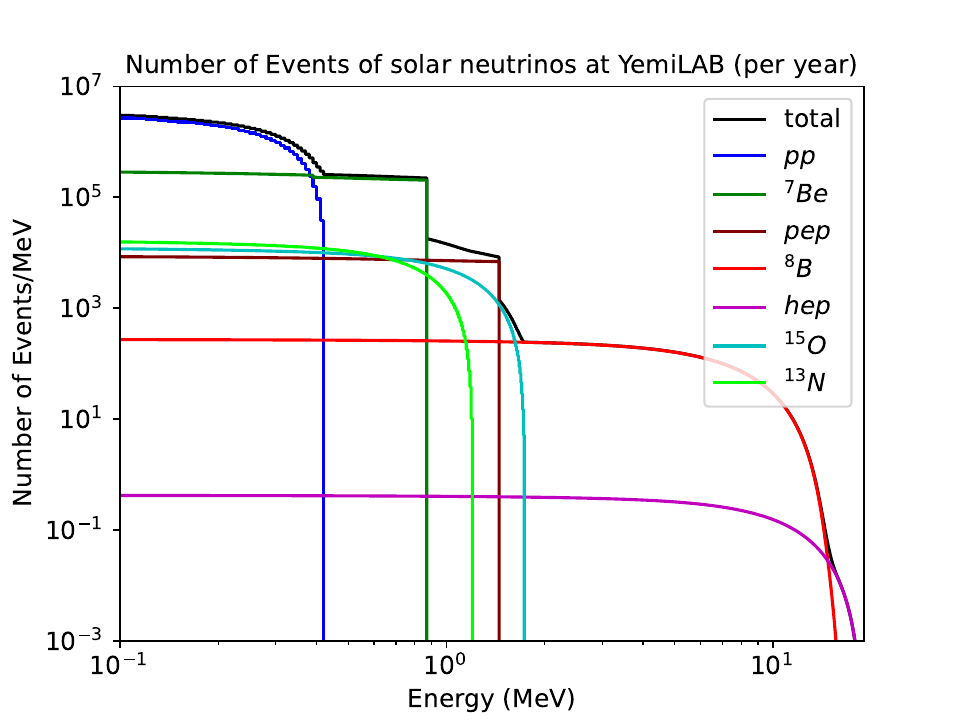}
\end{center}
\caption{\label{noe_solar1} The number of events per MeV per year as a function of the kinetic energy of the electron, assuming the best fit values of SK/SNO solar neutrino data ($\Delta m^2_{21} = 6.11 \times 10^{-5}$ eV$^2$ and $\sin^2\theta_{12} = 0.306$), is illustrated. The total number of events is represented by the black curve. The number of events for $pp$, $^7$Be, $pep$, $^8$B, $hep$, $^{15}$O, and $^{13}$N neutrinos are shown separately in dark blue, dark green, brown, red, magenta, light blue, and light green, respectively.}
\end{figure}

Figure~\ref{noe_solar1} depicts the annual number of events as a function of the electron kinetic energy in LSC at Yemilab, considering the best fit values of the SK/SNO data for $\Delta m^2_{21} =6.11\times 10^{-5}$  eV$^2$ and $\sin^2\theta_{12}=0.306$~\cite{nakajima2020recent}. The total number of events is represented by the black curve. Additionally, the number of events for the individual solar neutrino components, including $pp$, $^7$Be, $pep$, $^8$B, $hep$, $^{15}$O, and $^{13}$N, are illustrated separately using different colors: dark blue, dark green, brown, red, magenta, light blue, and light green, respectively. This plot provides an overview of the expected event rates for each solar neutrino source, allowing for a detailed analysis of their contributions to the total event count.

\begin{figure}[h]
\begin{center}
\includegraphics[width=0.52 \textwidth]{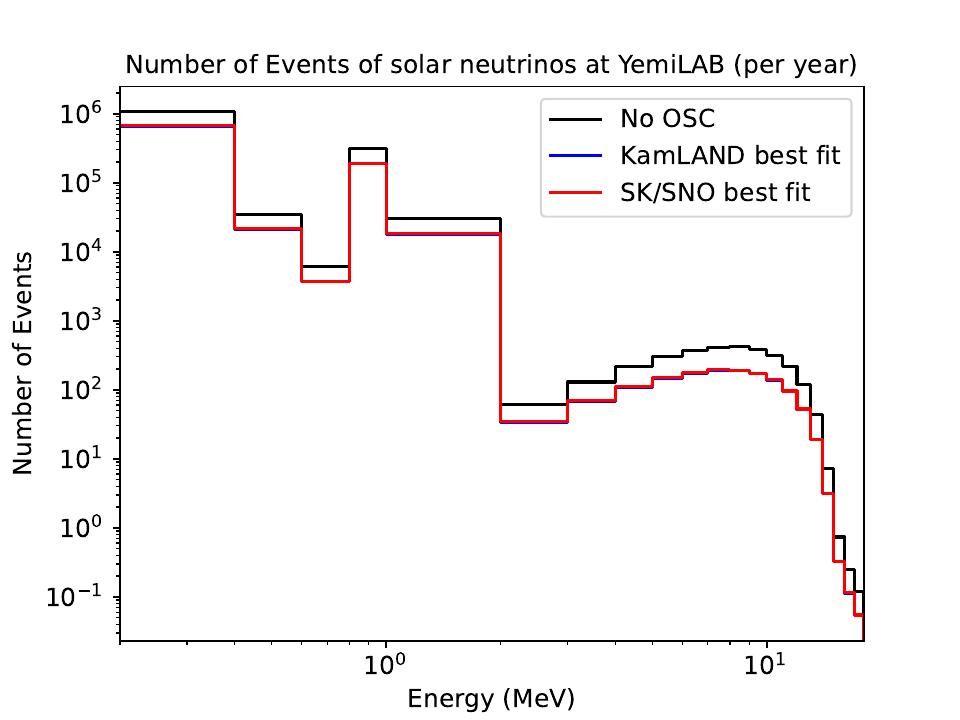}
\end{center}
\caption{\label{noe_solar2} Number of events per bin per year as a function of neutrino energy for different values of $\Delta m^2_{21}$ and $\theta_{12}$, corresponding to the best fit values of KamLAND data (blue curves) and the best fit values of SK/SNO solar neutrino data (red curves) and no oscillation (black curve). The number of events in each bin for the best fit values of KamLAND and solar data are close to each other, therefore, they coincide.}
\end{figure}

In Fig.~\ref{noe_solar2}, we have demonstrated the reconstructed number of events as a function of the neutrino energy, assuming no oscillation number of events (black curve), the best fit values of KamLAND data (blue curves) and the best fit values of SK/SNO solar neutrino data (red curves).  The number of events in each bin for the best fit values of KamLAND and SK/SNO solar data are close to each other, compared to the no oscillation case,
therefore they coincide for some bins.

\begin{figure}[h]
\begin{center}
\includegraphics[width=0.52 \textwidth]{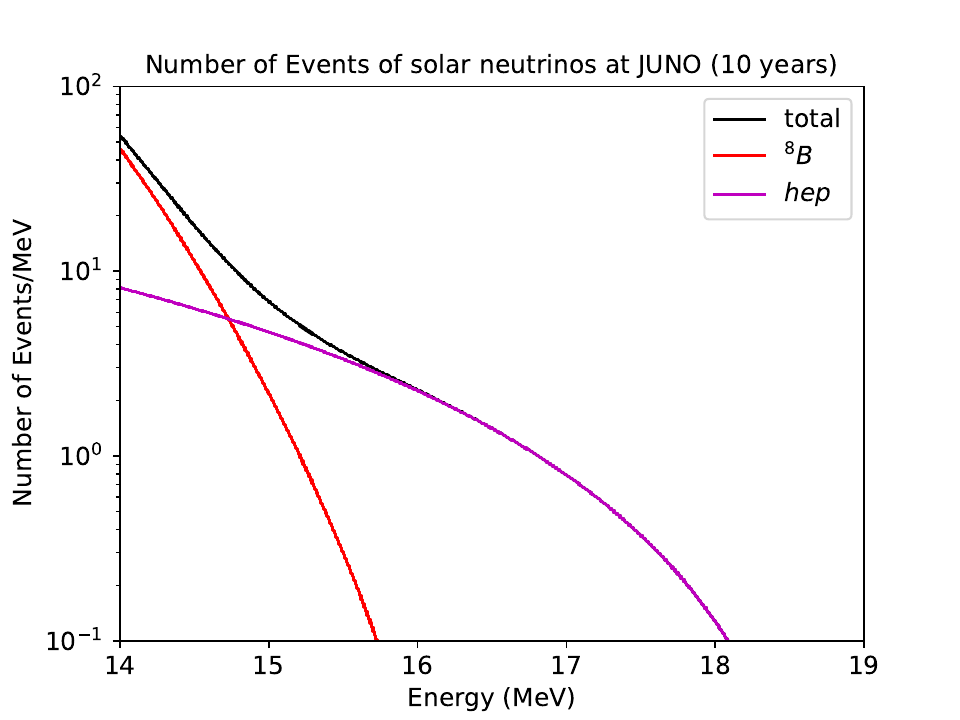}
\end{center}
\caption{\label{NOE_solar_JUNO} Number of events per MeV as a function of the kinetic energy of electron for the values of $\Delta m^2_{21}$ and $\theta_{12}$ fixed at the best fit values of SK/SNO solar data. We expect six events of $hep$ neutrinos and one event of $^8$B neutrinos at energies larger than 14.8 MeV after ten years of data taking at JUNO.}
\end{figure}

The number of events for the solar neutrino data as a function of the electron's kinetic energy is presented in the conceptual design of JUNO \cite{JUNO:2015zny}. Figure 6-3 of the JUNO conceptual design illustrates two scenarios for the background of low-energy solar data (below 1.8 MeV): one with high background and the other with a reduced lower background assumption based on ideal radiopurity conditions. As mentioned before, at lower energies, JUNO primarily detects $^7$Be neutrinos, with a small contribution from $pp$ neutrinos. For higher energies, JUNO is capable of detecting $^8$B neutrinos, as demonstrated in Figure 6-4 of the JUNO conceptual design \cite{JUNO:2015zny}. Neutron captures on the steel of the detector result in the production of 6 MeV and 8.5 MeV gamma rays, contributing to the external background and reducing the fiducial size of the detector by half. In our analysis on the other hand, 
we focus on the number of events at JUNO for energies larger than 14.5 MeV. Our results indicate that JUNO, with a 20 kton fiducial volume, is capable of detecting six events of $hep$ neutrinos with electron kinetic energies larger than 14.8 MeV after ten years of data taking. We anticipate observing approximately one event from $^8$B neutrinos within this energy range. It is worth noting that due to neutron captures and a potential reduction in the fiducial volume at these energies, the number of events may be lower. Figure~\ref{NOE_solar_JUNO} displays the number of events as a function of kinetic energy for energies larger than 14 MeV. Notably, for energies above 14.8 MeV, the number of $hep$ neutrino events exceeds that of $^8$B neutrinos.

\subsection{Sensitivities on the solar neutrino parameters}
\label{sec:sesnsitivity-results}

\begin{figure}[h]
\begin{center}
\includegraphics[width=0.49 \textwidth]{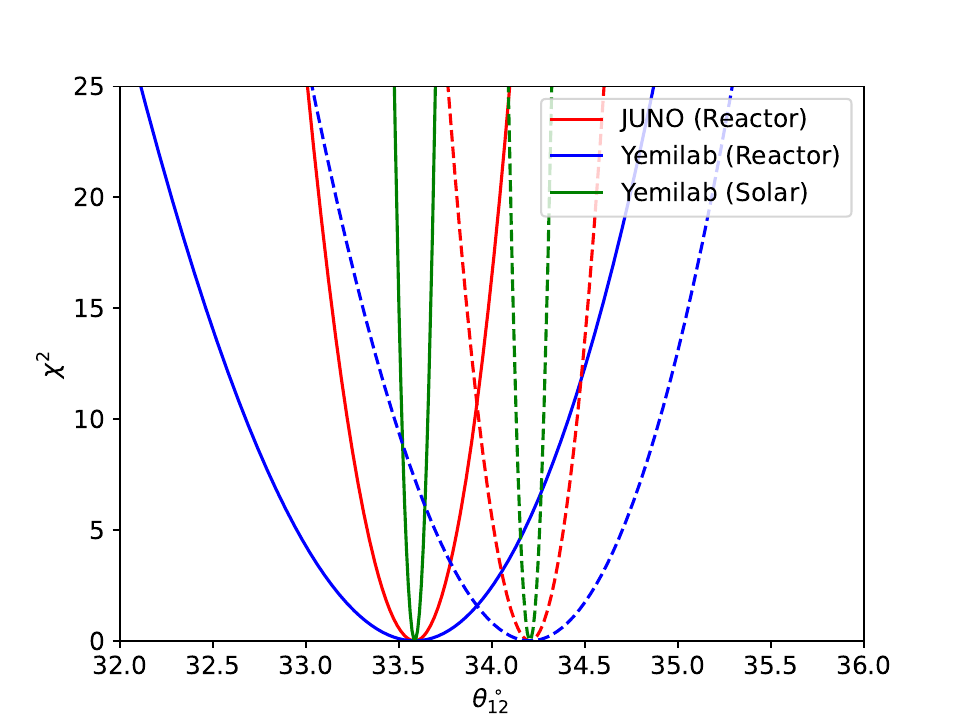}
\includegraphics[width=0.49 \textwidth]{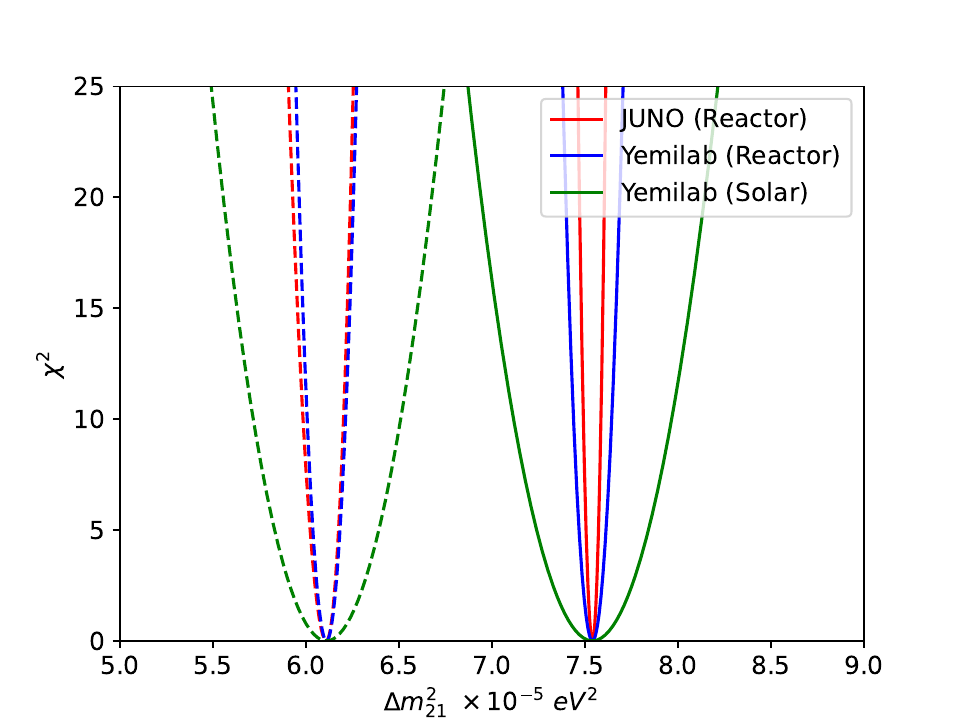}
\end{center}
\caption{\label{constraints_1d} The constraints on solar neutrino oscillation parameters $\theta_{12}$ and $\Delta m^2_{21}$ using detection of solar neutrinos in LSC at Yemilab and detection of reactor neutrinos both in LSC at Yemilab and JUNO for 10 years. The dashed curves are plotted based on the assumption that the best fit values of SK/SNO solar data are the true value, while the solid curves are plotted assuming the best fit values of KamLAND as the true value. The blue, red, and green curves correspond to the Yemilab reactor, JUNO, and Yemilab solar data, respectively.
As it is demonstrated, solar neutrino in LSC at Yemilab determines the value of $\theta_{12}$ with the highest precision. JUNO will determine the value of $\Delta m^2_{21}$ with the highest precision.}
\end{figure}

In Fig.~\ref{constraints_1d}, we 
show the sensitivities 
on solar neutrino oscillation parameters $\theta_{12}$ and $\Delta m^2_{21}$ from solar neutrino detection in LSC at Yemilab and reactor neutrino detection both in LSC at Yemilab and JUNO for 10 years, while fixing oscillation parameters $\delta_{\rm CP} = 0$, $\Delta m^2 _{31}$ from nu-fit \cite{Esteban:2020cvm}, and $\theta_{13}$ as the best fit value of Daya Bay \cite{DayaBay:2018yms}, assuming normal mass ordering.
As it is shown, 
we expect the solar neutrino data in LSC at Yemilab can determine the value of $\theta_{12}$ with the highest precision, 
albeit with the smallest size of the fiducial volume among the considerations listed in Table~\ref{table:fiducial}.
The main reason 
is its low energy threshold  
along with 
detection of $pp$ neutrinos and $^7$Be neutrinos with high statistics with the order of several hundred thousand events per year.

On the other hand, the detection of the reactor neutrinos has a better sensitivity to determine the value of $\Delta m^2_{21}$. JUNO detector with 
a larger fiducial volume, 
more powerful reactors and closer baseline
can provide a better sensitivity than LSC at Yemilab if $\Delta m^2 _{21}$ is closer to the KamLAND best fit value.

We demonstrate the result assuming two cases: i) the true values of the solar neutrino oscillation parameters are the best fit values of 
the reactor neutrino results from 
KamLAND, ii) the best fit values of the SK/SNO solar data as the true values for solar neutrino oscillation parameters. 
As it is demonstrated in the figure, for larger values of $\Delta m^2_{21}$ closer to the best fit value of KamLAND, JUNO has approximately twice better sensitivity to determine the value of $\Delta m^2_{21}$, however, if $\Delta m^2_{21}$ is smaller and it is close to the best fit value of the SK/SNO solar data, the sensitivity of JUNO is comparable with LSC at Yemilab. Later, we will show the potential of these experiments for simultaneous measurement of $\Delta m^2_{21}$ and $\theta_{12}$. For smaller values of $\Delta m^2_{21}$, closer to the best fit values of SK/SNO solar data, LSC at Yemilab reactor neutrino will have a better sensitivity to determine $\Delta m^2_{21}$ more precisely than JUNO, while for larger values of $\Delta m^2_{21}$, close to the best fit values of KamLAND, JUNO has a better sensitivity to determine $\Delta m^2_{21}$, in the presence of $\theta_{12}$ or marginalizing over $\theta_{12}$.

\begin{figure}[h]
\begin{center}
\includegraphics[width=0.49 \textwidth]{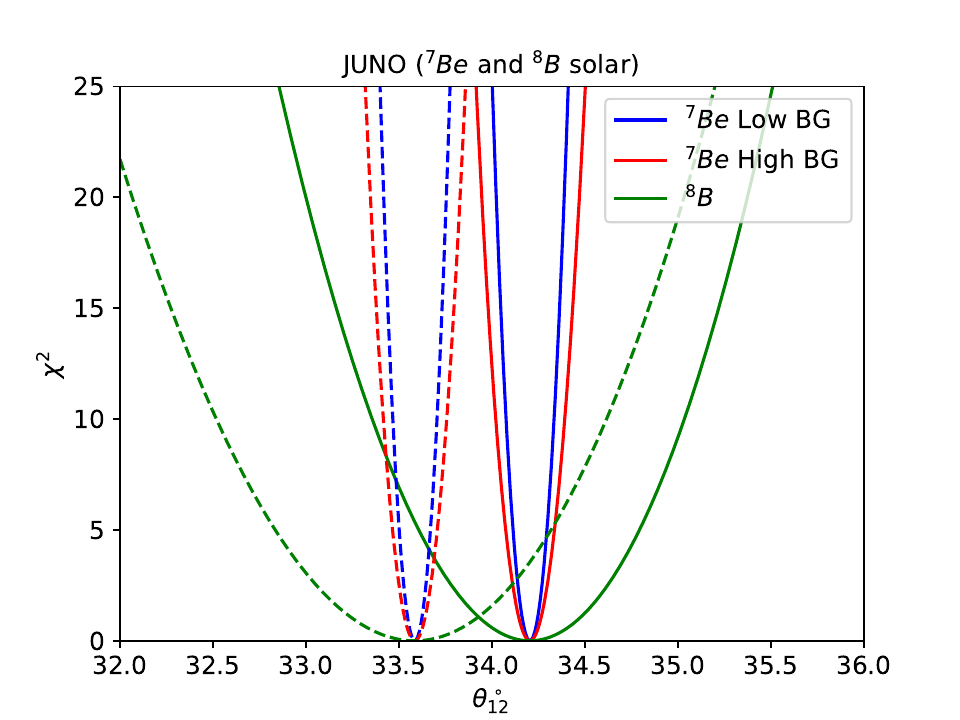}
\includegraphics[width=0.49 \textwidth]{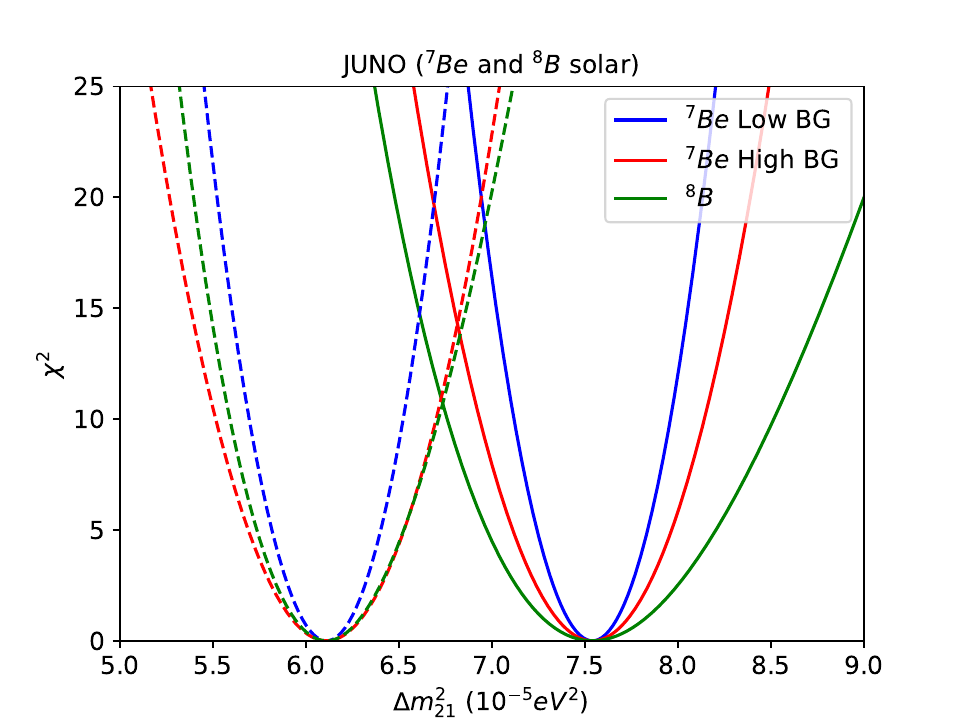}
\end{center}
\caption{\label{constraints_1d_juno} 
The constraints on solar neutrino oscillation parameters $\theta_{12}$ and $\Delta m^2_{21}$ with detection of solar neutrinos at JUNO, assuming ten years of data collection. The dashed curves are plotted based on the assumption that the best fit values of SK/SNO solar data is the true value, while the solid curves are plotted assuming the best fit values of KamLAND as the true value.
The green curves correspond to the $^8$B solar neutrinos detected at JUNO. The blue and red curves are plotted assuming low background and high background for the $^7$Be neutrinos, respectively.
As observed, the solar neutrino detection at JUNO exhibits lower sensitivity in measuring $\theta_{12}$ and $\Delta m^2_{21}$ compared to the reactor neutrino detection at JUNO and Yemilab, as well as the solar neutrino detection at Yemilab.
}
\end{figure}

In Fig.~\ref{constraints_1d_juno}, we 
show the sensitivities of JUNO on the 
solar neutrino oscillation parameters $\theta_{12}$ and $\Delta m^2_{21}$
from the detection of the solar neutrinos.
Notice that, due to the large amount of $^{14}$C background below 0.2 MeV, only the electron with larger values of kinetic energy can be detected. For energies larger than 0.2 MeV, there are also other sources of background
such as $^{210}$Bi and $^{11}$C so that JUNO can only detect the $^7$Be neutrino at low energies. In this case, the background is two times larger than the signal \cite{JUNO:2015zny}.
For the case of ideal radiopurity background,  only $^7$Be neutrinos are detectable and the number of events of the signal is about one order of magnitude larger than the background. Also, the other sources of low energy solar neutrinos will be much smaller than the background. 
For higher energies, we have a large amount of $^{10}$C and $^{11}$C background, which are one to two orders of magnitude larger than $^8$B neutrino events, for energies smaller than 3.5 MeV.

Comparing Figs.~\ref{constraints_1d} and \ref{constraints_1d_juno}, we observe that under the assumption of low background, the constraint on $\Delta m^2_{21}$ derived from the $^7$Be solar data of JUNO is comparable to, but slightly weaker, than that obtained from the solar data of LSC at Yemilab. Note that solar data from both LSC at Yemilab and JUNO can result in weaker constraints on $\Delta m^2_{21}$ compared to reactor data. It is worth noting that the constraint obtained using $^8$B solar data at JUNO is significantly weaker than reactor data of JUNO and Yemilab, and solar data at Yemilab constraints. In obtaining the constraints on $\Delta m^2 _{21}$, we have fixed the value of $\theta_{12}$ as the best fit values from KamLAND and SK/SNO.

In the determination of $\theta_{12}$, the $^8$B solar data at JUNO is observed to exhibit less sensitivity compared to the $^7$Be data. However, when considering the $^7$Be solar data at JUNO, a constraint on $\theta_{12}$ is obtained that is comparable, albeit slightly less sensitive, to the solar data at LSC at Yemilab. It is important to note that the flux uncertainties have not been accounted for in our analysis. Including these uncertainties would lead to weakened constraints, especially for the solar data at JUNO, as $^7$Be and $^8$B neutrinos are associated with significant theoretical flux uncertainties. In our analysis for determining $\theta_{12}$, we have fixed the value of $\Delta m^2_{21}$ as the best fit values from KamLAND and SK/SNO. 

Nevertheless, the detection of solar neutrinos at JUNO will play key roles in determining the individual fluxes of $^7$Be and $^8$B as well as their ratio.
Moreover, the detection of $hep$ neutrinos and their flux measurement will be another astonishing potential of JUNO as it is demonstrated in Fig.~\ref{NOE_solar_JUNO}. 
We expect the
combination of JUNO and LSC at Yemilab solar neutrinos 
can open a new window to study the structure of the Sun by precise measurements of all the solar neutrino fluxes and their ratios.

It is important to emphasize that the primary goal of JUNO is the determination of the neutrino mass ordering, and its baseline is optimized for this purpose. Additionally, if the value of $\Delta m^2_{21}$ is close to the best fit value determined by KamLAND, JUNO can provide a highly precise measurement of $\Delta m^2_{21}$. However, for smaller values of $\Delta m^2_{21}$, a longer baseline is more effective. It is worth noting that JUNO benefits from a ten times larger detector, reactor power that is one and a half times greater, and a flux that is $(65~\text{km}/52.5~\text{km})^2$ higher due to the closer baseline.

\begin{figure}[h]
\begin{center}
\includegraphics[width=0.6 \textwidth]{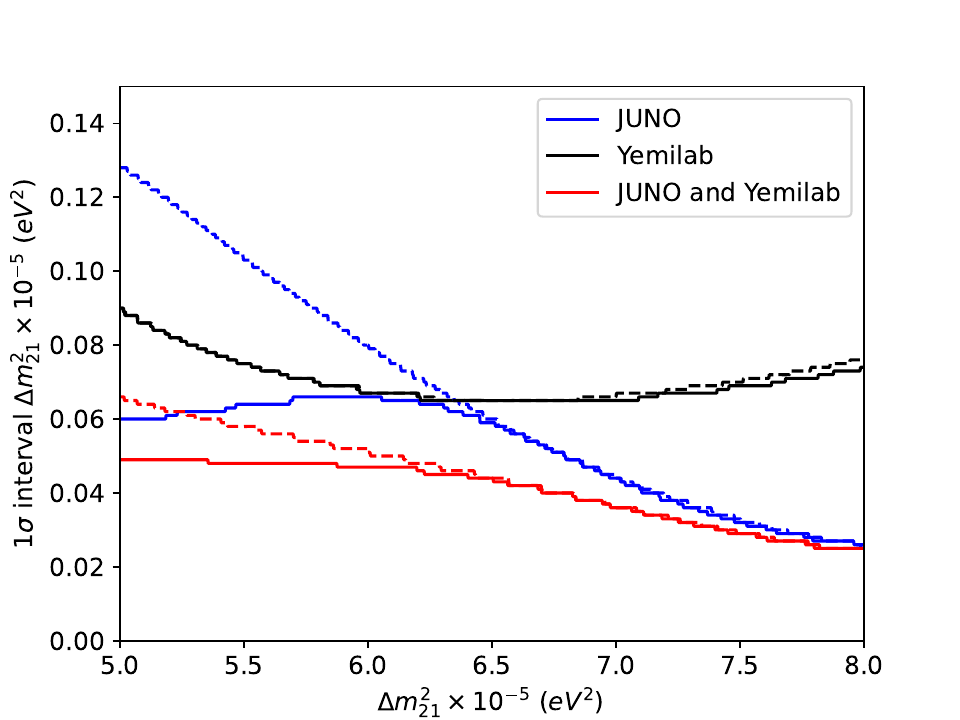}
\end{center}
\caption{\label{interval} The one sigma interval that JUNO and LSC at Yemilab 
can achieve for different values of $\Delta m^2_{21}$ after ten years of reactor neutrino data taking. As shown, for smaller values of $\Delta m^2_{21}$, JUNO has less sensitivity to determine $\Delta m^2_{21}$, and its precision will be closer to that of Yemilab. However, JUNO exhibits better sensitivity for larger values of $\Delta m^2_{21}$.  Dashed curves are plotted when marginalizing over $\theta_{12}$. The solid curves are plotted setting all oscillation parameters to their best fit values \cite{nakajima2020recent, Esteban:2020cvm, DayaBay:2018yms}.} 
\end{figure}

In Fig.~\ref{interval}, we illustrate the sensitivities of JUNO and LSC at Yemilab in determining the value of $\Delta m^2_{21}$
from detecting the reactor neutrinos. The $1\sigma$ confidence level intervals are shown for different values of $\Delta m^2_{21}$.  The solid line represents a fixed value of $\theta_{12}$ to the best fit values of KamLAND,
while the dashed line represents the $1\sigma$ interval obtained through marginalization over $\theta_{12}$. As demonstrated, for larger values of $\Delta m^2_{21}$, JUNO exhibits two times better sensitivity in determining $\Delta m^2_{21}$ compared to LSC at Yemilab. On the other hand, for smaller values of $\Delta m^2_{21}$, LSC at Yemilab shows a better sensitivity. It should be noted that despite the similar flux shape and cross sections of neutrinos from the reactors, the different baselines result in distinct neutrino oscillation probability patterns and, consequently, different patterns of events per energy bin.   Notably, when considering $\Delta m^2_{21}<6.4\times10^{-5}$ and assuming marginalization over $\theta_{12}$, the reactor neutrino detection in LSC at Yemilab demonstrates superior sensitivity for measuring $\Delta m^2_{21}$. Moreover, the determination of $\Delta m^2_{21}$ using reactor neutrinos with two different baselines helps in reducing the systematic uncertainties associated with the reactor neutrino flux or other sources of systematic uncertainties
of the quantities such as $\theta_{12}$ and backgrounds. Notice that the result from Yemilab is robust when considering the marginalization of $\theta_{12}$ since the minimum of $P_{ee}$ or the maximum flavor conversion occurs within the energy range where the largest number of events is observed (Fig.~\ref{pee_reactor}). 
Therefore, if the true value of $\Delta m^2_{21}$ is $6.11 \times 10^{-5}$ eV$^2$, we will have a better sensitivity to both $\theta_{12}$ and $\Delta m^2_{21}$. However, this is not the case for JUNO, as the maximum flavor conversion occurs around the energy range of $2-3$ MeV where the number of events is lower around 30$\%$. As a result, a strong anti-correlation is present between $\theta_{12}$ and $\Delta m^2_{21}$ in the case of JUNO.


\begin{figure}[h] 
\begin{center}
\includegraphics[width=0.49 \textwidth]{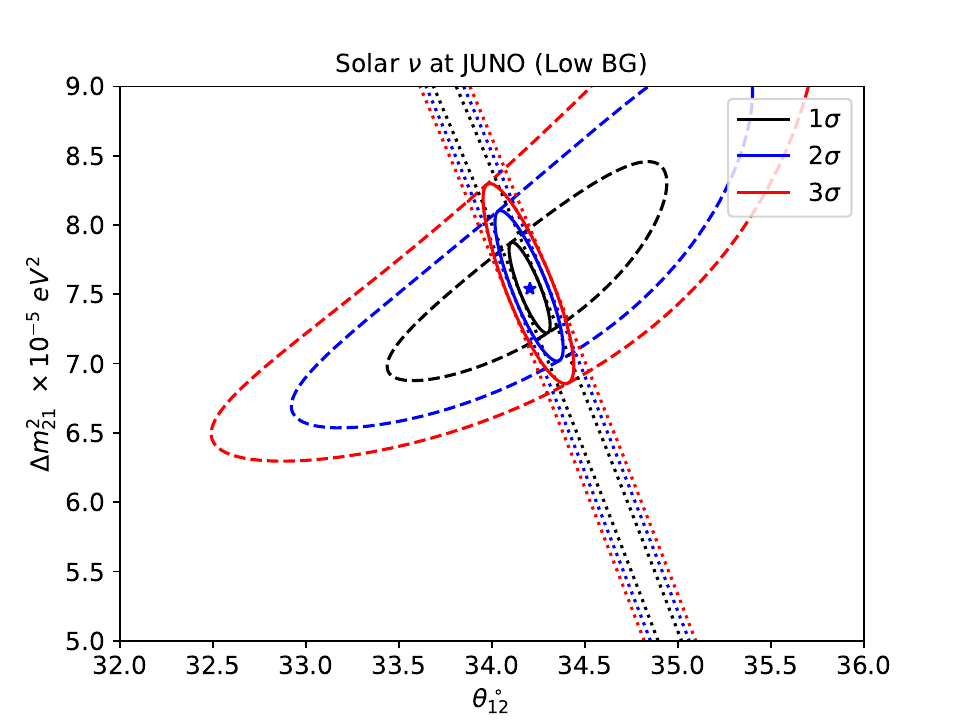}
\includegraphics[width=0.49 \textwidth]{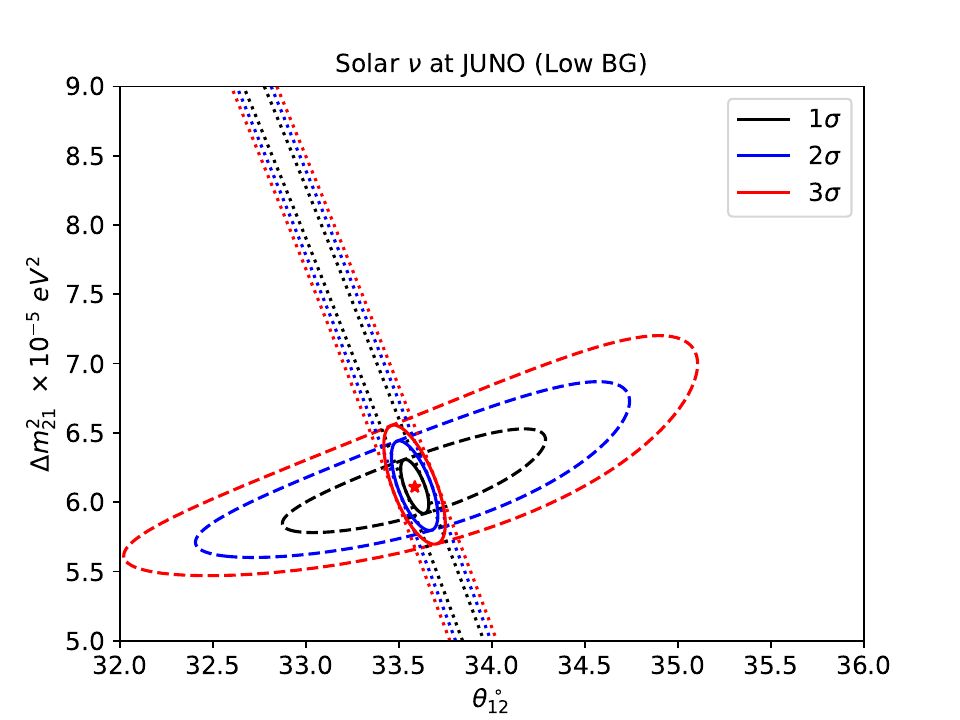}
\includegraphics[width=0.49 \textwidth]{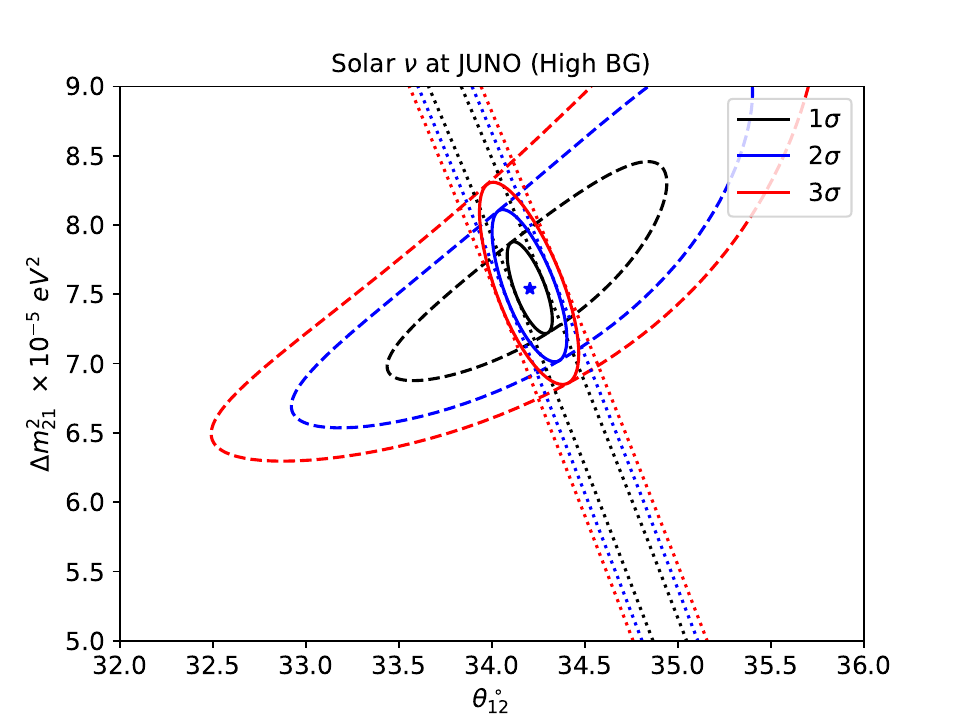}
\includegraphics[width=0.49 \textwidth]{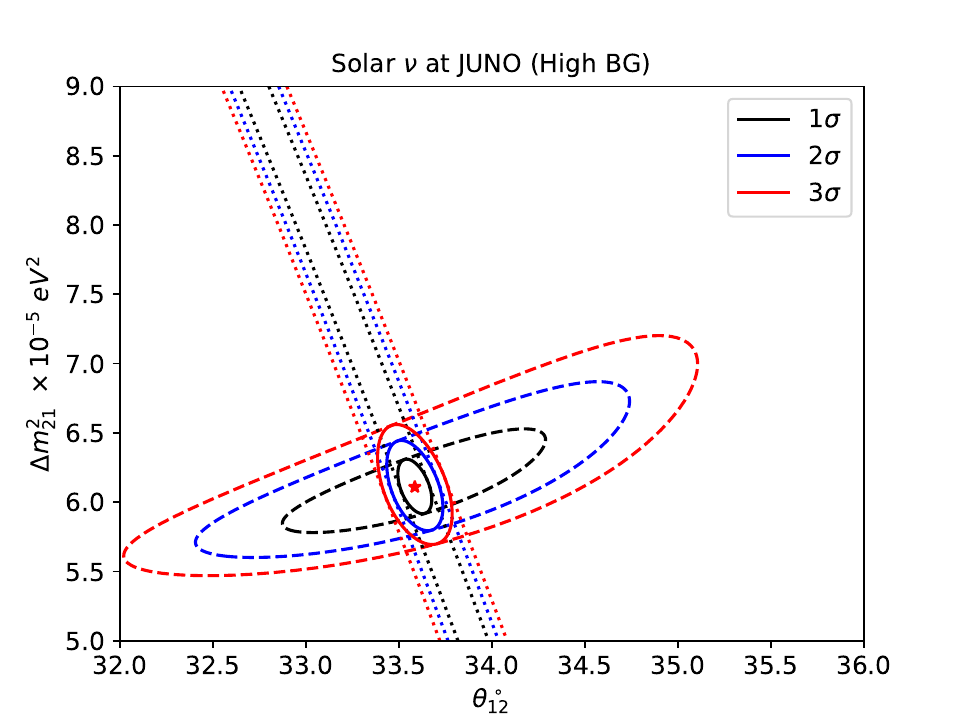}
\includegraphics[width=0.49 \textwidth]{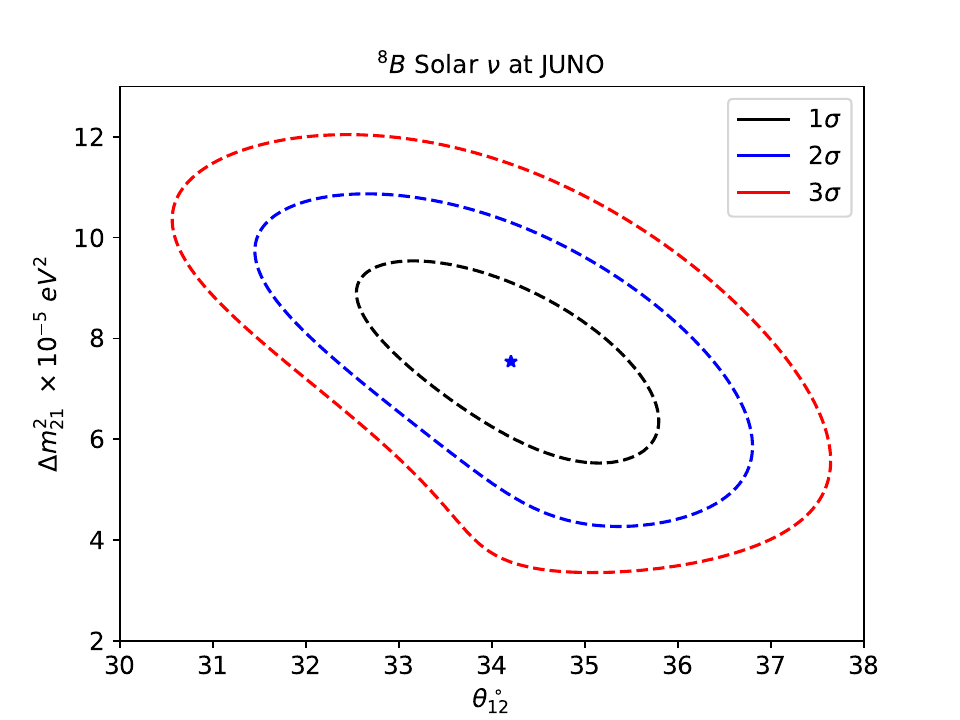}
\includegraphics[width=0.49 \textwidth]{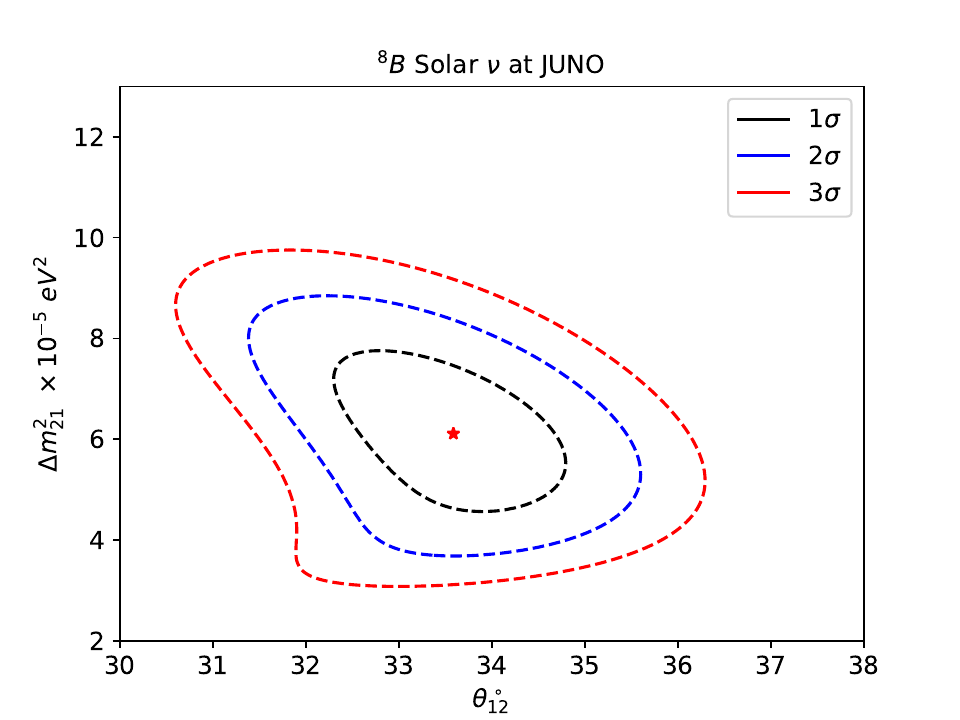}
\end{center}
\caption{\label{constraints_2d_JUNO_solar} Simultaneous measurement of  $\Delta m^2_{21}$ and $\theta_{12}$ using solar neutrino detection at JUNO, assuming ten years of data collection, with low (upper panels) and high (middle panels) background levels considering the best fit values of KamLAND (left panels) and SK/SNO (right panels). The dashed and dotted curves correspond to the $^8$B and $^7$Be neutrinos, respectively. The solid curves represent the combination of $^8$B and $^7$Be neutrinos. The bottom panel indicates the constraint obtained considering only $^8$B including $3.5\%$ flux uncertainty. As can be observed, including the flux uncertainty will wash out the sensitivity.}
\label{sol11}
\end{figure}

We have also studied the potential of LSC at Yemilab to determine the mass ordering, which is sensitive to the energy resolution of the detector. After ten years of data taking, LSC at Yemilab will determine the mass ordering with 1$\sigma$ C.L. for $5\%\sqrt{E}$, $4\%\sqrt{E}$ energy resolution and 2$\sigma$ C.L. for $3\%\sqrt{E}$ energy resolution.
On the other hand, the sensitivity of more dedicated experiment JUNO is beyond 5$\sigma $ C.L. after five years of data taking with the energy resolution of the detector ranging from $3\%\sqrt{E}$ to $3.5\%\sqrt{E}$. Note that, the true value of the solar oscillation parameters $\Delta m^2_{21}$ and $\theta_{12}$ can affect determination of mass ordering by JUNO \cite{Forero:2021lax}.

Figure \ref{sol11}  shows the  simultaneous measurement of  $\Delta m^2_{21}$ and $\theta_{12}$ using solar neutrino detection at JUNO for 10 years.  The figure presents two distinct scenarios: a low background case (upper panels) and a high background case (middle panels). In the left panels, we assume the fixed value of $\theta_{12}$ to be the best fit value obtained from KamLAND, while in the right panels, we fix it as the best fit value derived from SK/SNO. The dashed curves correspond to the $^8$B solar data, while the dotted curves correspond to the $^7$Be solar data. The solid contours correspond to the combined data of $^7$Be and $^8$B solar data.
As can be observed, the $^8$B data exhibits sensitivity to the simultaneous measurement of $\Delta m^2_{21}$ and $\theta_{12}$ due to the significant impact of the resonance effect within its energy range.  However, as previously mentioned, the oscillation probability of $^7$Be monochromatic solar neutrinos shows an anti-correlated dependence on $\theta_{12}$ and $\Delta m^2_{21}$. This behavior can be observed from Eq.~(\ref{costheta}), particularly for small values of $\epsilon_{12}$, where $\cos 2 \theta ^m _{12} \sim \cos 2 \theta_{12} - \frac{A_{CC}}{\Delta m^2 _{21}}$, aligning with our expectations for an anti-correlated relationship. By combining both the $^8$B and $^7$Be solar data obtained at JUNO, a more stringent constraint can be achieved. Notice that when considering the sensitivity of solely $^8$B solar data, its constraints are considerably weaker in comparison to JUNO reactor data, as well as solar and reactor data from LSC at Yemilab. Thus, detecting only  $^8$B  neutrinos will not lead to a significant constraint. In the top and middle panels,  we have not included the flux uncertainty of the solar neutrinos as well as the other systematics. 
However, it is worth mentioning that in the case of $pp$ neutrinos which could be detected at LSC, the flux uncertainty is very small, approximately $0.6\%$ \cite{Vinyoles:2016djt}. 
This robustly supports our result for LSC at Yemilab, as including the $pp$ flux uncertainty would not significantly change our conclusion that LSC will provide the best sensitivity for measuring $\theta_{12}$.~\footnote{Note, however, that a more dedicated analysis of estimating other possible systematic uncertainties at LSC is required.} On the other hand, the flux uncertainty for $^7$Be and $^8$B neutrinos is $6\%$ and $12\%$, respectively, implying that including these uncertainties will weaken the obtained constraint on JUNO solar neutrinos. In the bottom panels, we are showing the simultaneous measurement of  $\Delta m^2_{21}$ and $\theta_{12}$ using $^8$B solar neutrino detection at JUNO with including $3.5\%$ flux uncertainty. We have chosen this value for flux uncertainty to compare our results with that of Ref.~\cite{JUNO:2020hqc} and we find that our results are in agreement with theirs. 
Note that 14 different systematics is included in the analysis performed in Ref.~\cite{JUNO:2020hqc},
while only the flux uncertainty systematic is included in the lower panels of Fig.~\ref{sol11}.
As observed, including the flux uncertainty can significantly relax the constraint. In the same way, including other systematics can wash out the sensitivity of JUNO to the $^8$B solar neutrino detection.

\begin{figure}[h]
\begin{center}
\includegraphics[width=0.45 \textwidth]{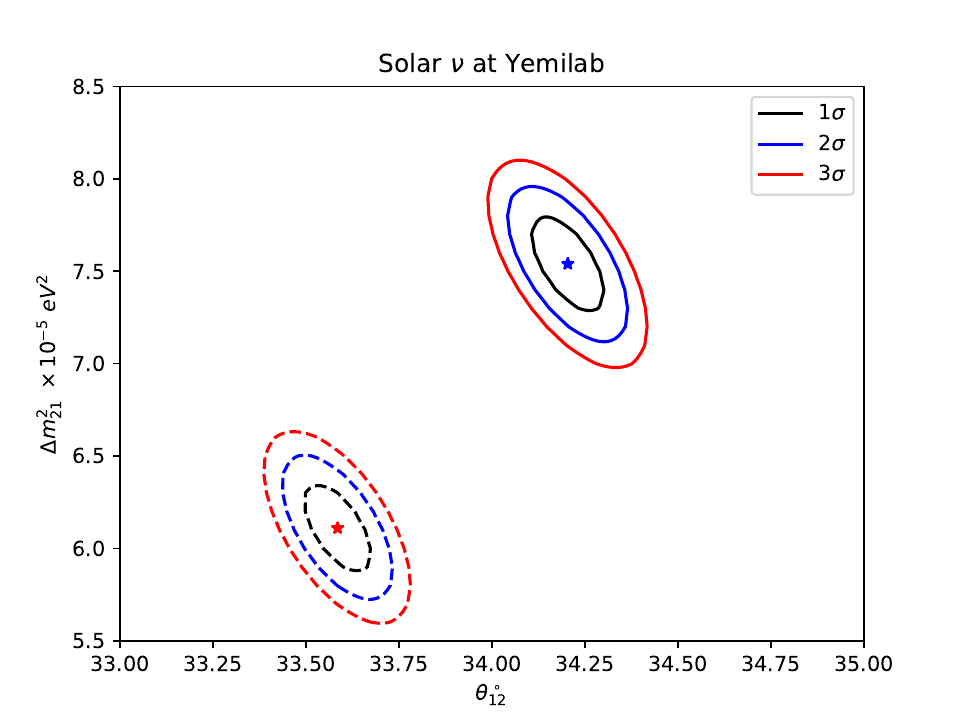}
\includegraphics[width=0.45 \textwidth]{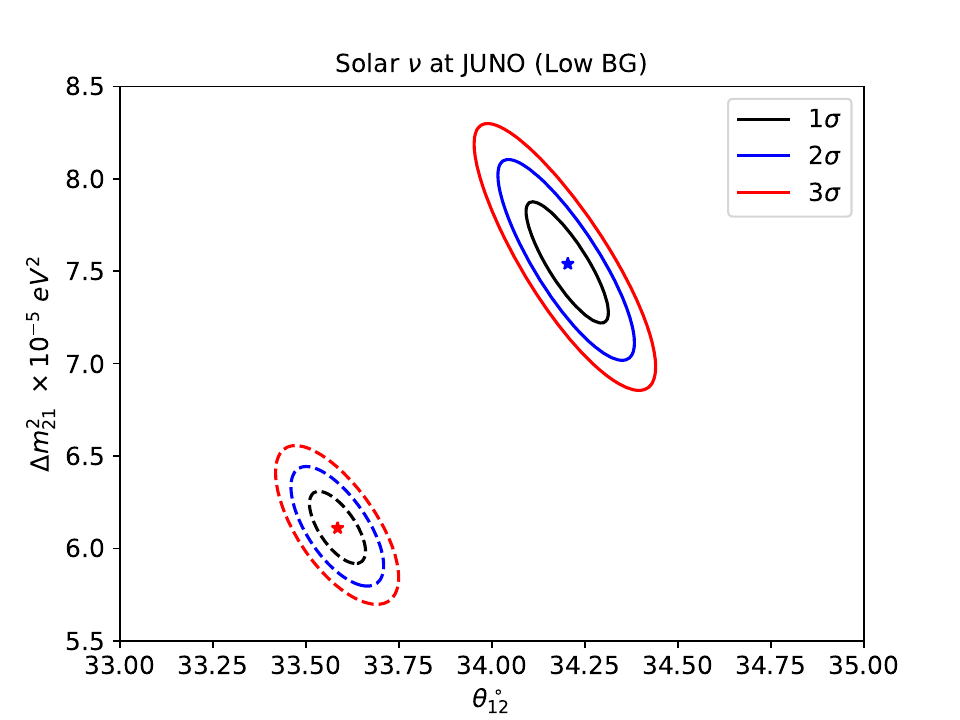}
\includegraphics[width=0.45 \textwidth]{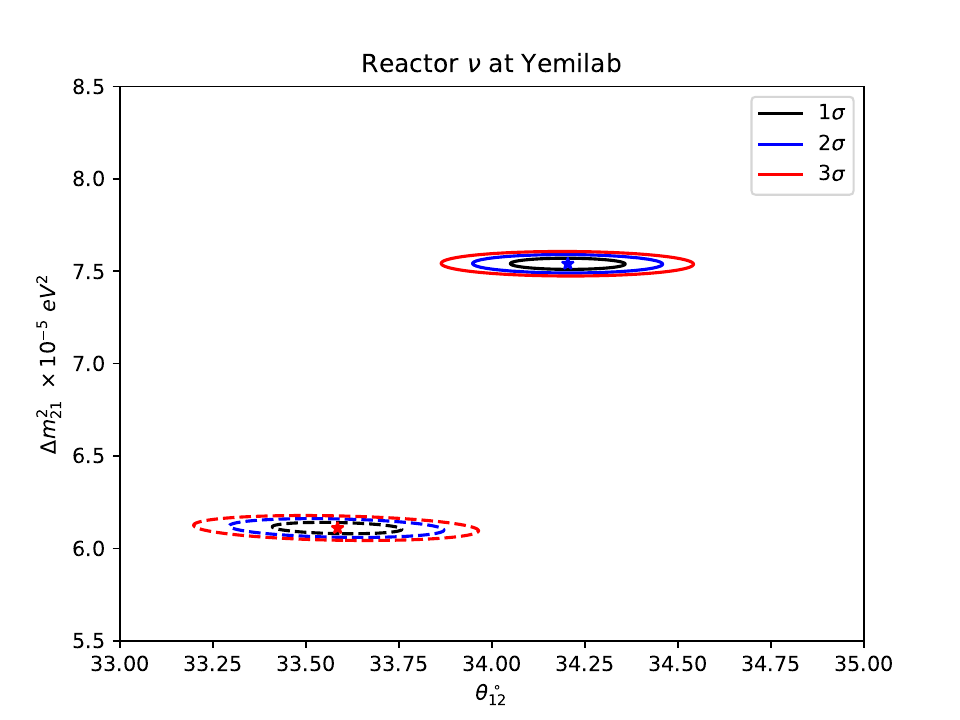}
\includegraphics[width=0.45 \textwidth]{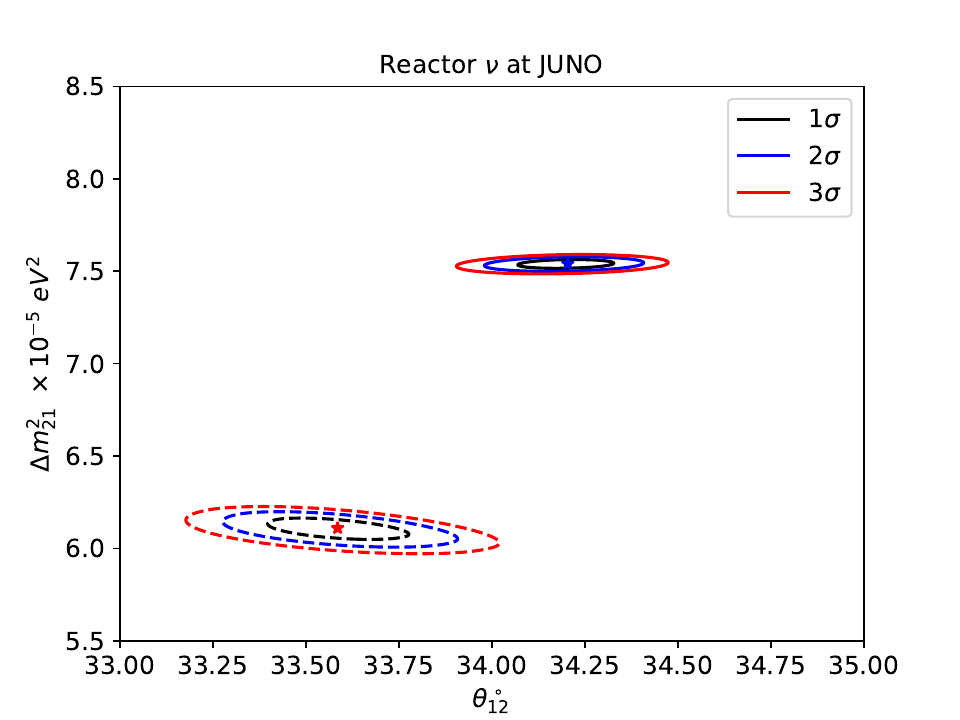}
\includegraphics[width=0.45 \textwidth]{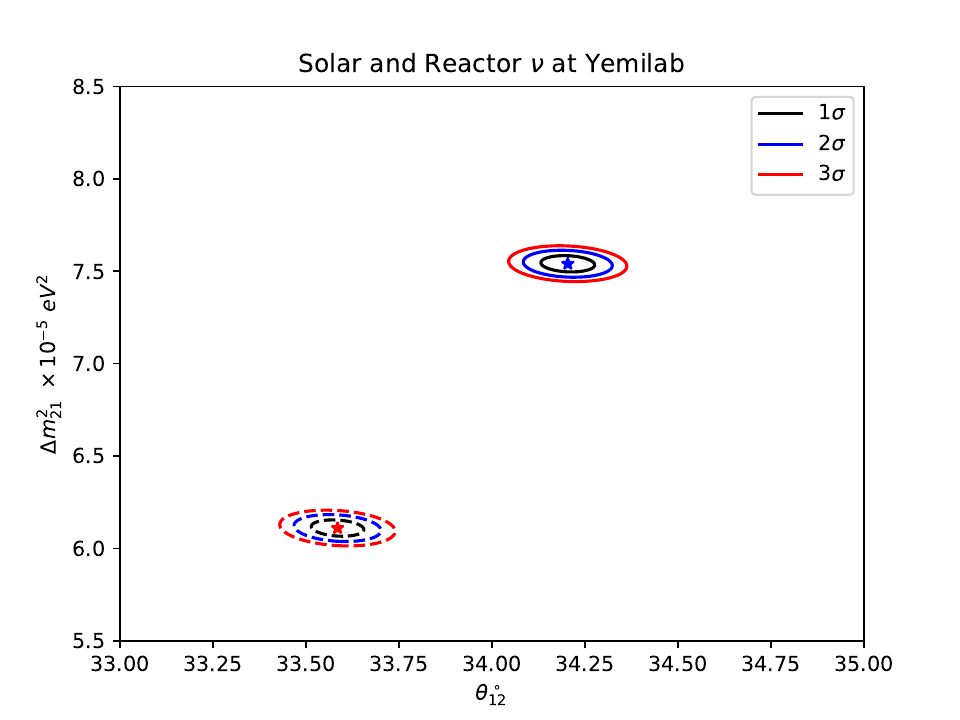}
\includegraphics[width=0.45 \textwidth]{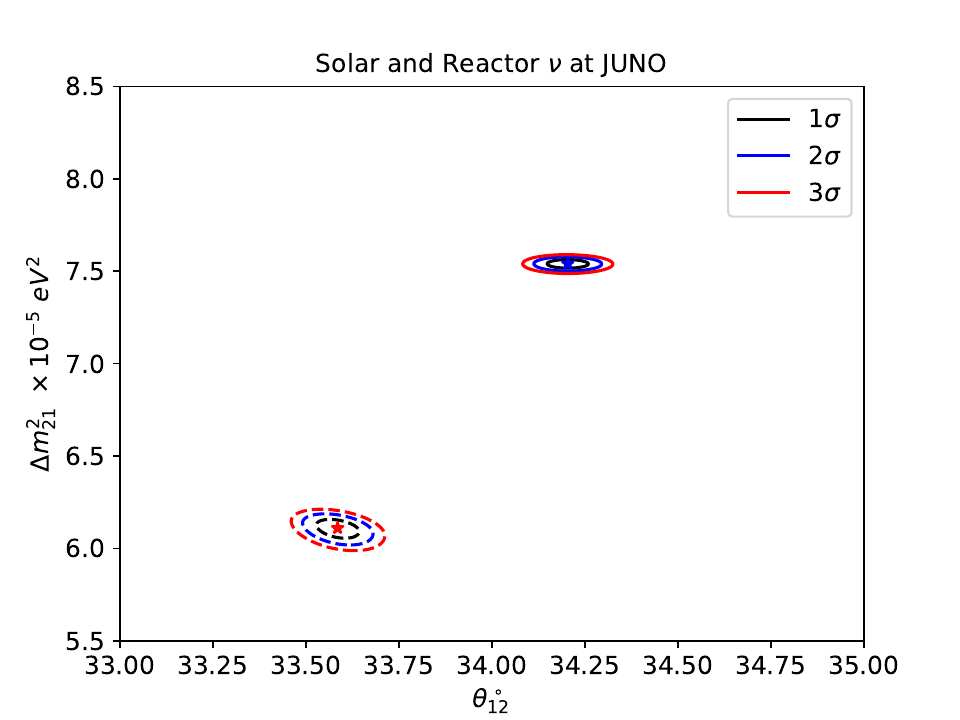}
\includegraphics[width=0.45 \textwidth]{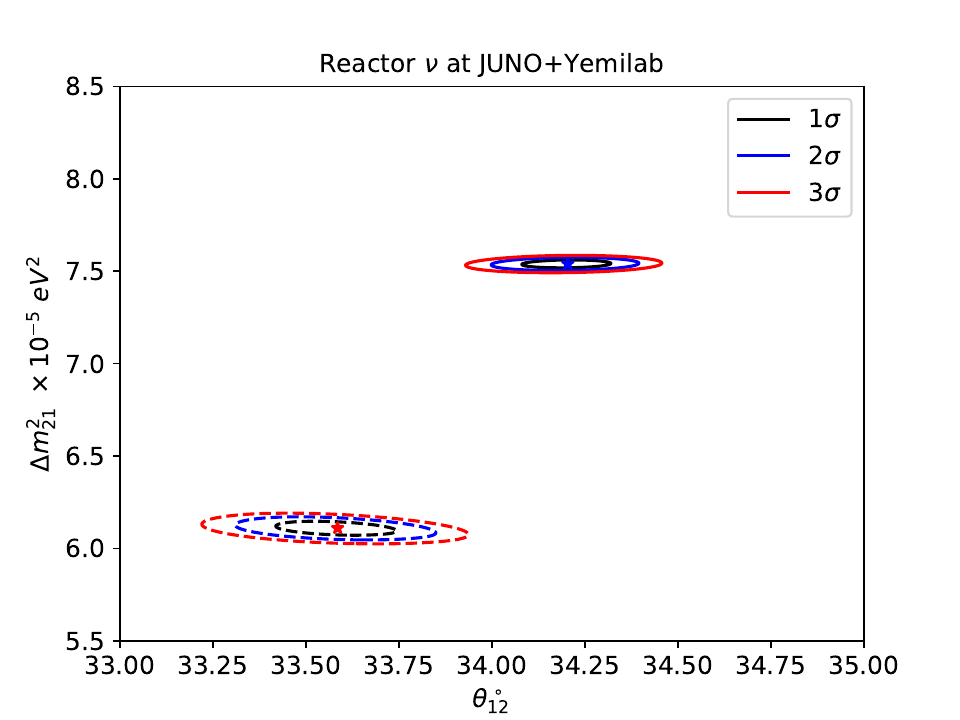}
\includegraphics[width=0.45 \textwidth]{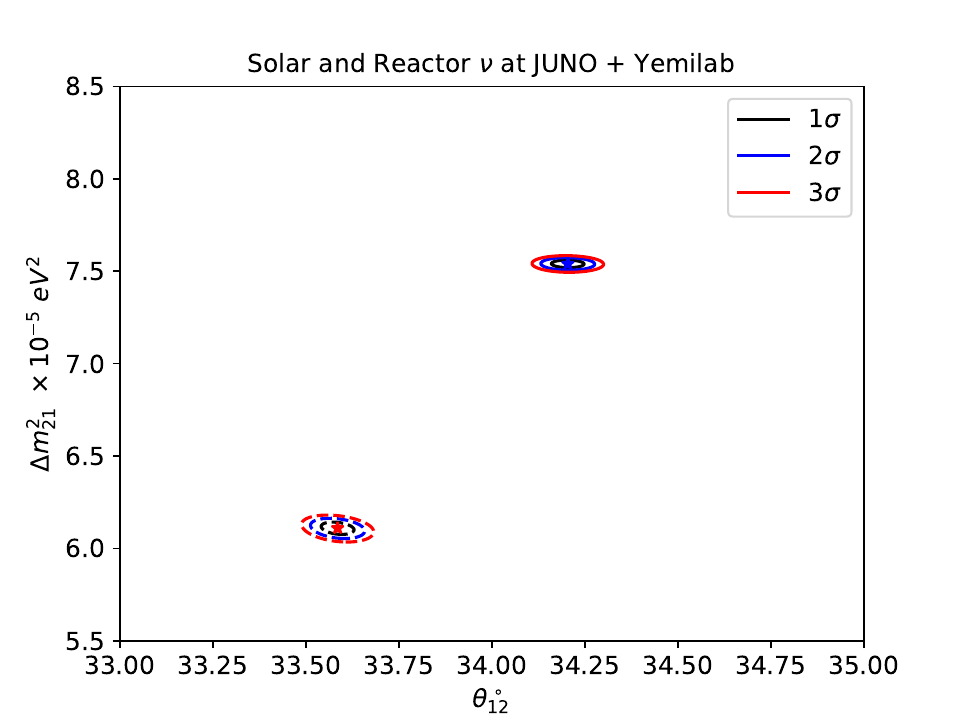}
\end{center}
\caption{\label{t12_dm21} Simultaneous measurement of $\theta_{12}$ and $\Delta m^2_{21}$ can be achieved through the detection of solar and reactor neutrinos at Yemilab, as well as the detection of reactor neutrinos at JUNO. Additionally, combining the data from both experiments can improve the sensitivity.
} 
\end{figure}

Figure~\ref{t12_dm21} demonstrates the sensitivities of LSC at Yemilab and JUNO to simultaneous measurement of $\theta_{12}$ and $\Delta m^2_{21}$. For the true values of $\theta_{12}$ and $\Delta m^2_{21}$, we have assumed two cases: the best fit values of KamLAND and that of SK/SNO solar data, as stated earlier.
In the top panels, we show the sensitivity of LSC at Yemilab (left) and JUNO (right) using ten years of solar data only. We have included the low background for JUNO here which leads to a more stringent constraint compared to the high background case.
In the second row, the sensitivity at LSC at Yemilab (right) and JUNO (left) using ten years of reactor data is demonstrated. 
In the panel on the right side of the third row, the expected sensitivity of LSC at Yemilab is displayed after combining both solar and reactor data. The panel on the left side of the third row represents the combined solar and reactor data at JUNO. It can be observed that the results of combining solar and reactor data at LSC at Yemilab are comparable to those of JUNO. However, it should be noted that the inclusion of flux uncertainty and high background for JUNO has the potential to weaken the obtained constraint.

Noticeably, solar neutrino detection by LSC at Yemilab can determine the value of $\theta_{12}$ with the highest precision.
For the simultaneous measurement of $\theta_{12}$ and $\Delta m^2_{21}$, LSC at Yemilab with the detection of both solar and reactor neutrinos will be comparable with or better than JUNO with reactor neutrinos only.
Note that in the case of using reactor neutrinos only, LSC at Yemilab shows a comparable sensitivity on $\Delta m^2_{21}$ for the SK/SNO best fit values ($\Delta m^2_{21} =6.11\times 10^{-5}$  eV$^2$ and $\sin^2\theta_{12}=0.306$ \cite{nakajima2020recent}), while for the other case, JUNO has a better sensitivity to determine the value of $\Delta m^2_{21}$. 
In the bottom panels, we show the combined sensitivity: with the reactor neutrino data only (left) and both the solar and reactor neutrino data from LSC at Yemilab and reactor data from JUNO (right). 
Remarkably, we expect that the combination of LSC in Yemilab and JUNO can determine the parameters $\theta_{12}$ and $\Delta m^2_{21}$ with a sub-percent precision level. 
Comparing the bottom-left and right panels, we can see the important contribution of solar neutrino detection in LSC at Yemilab in  improving precision.

\section{Conclusions}
\label{sec:conclusions}

We have explored the sensitivities of LSC at Yemilab and JUNO to solar neutrino oscillation parameters $\theta_{12}$ and $\Delta m^2 _{21}$.  
In our analysis, we have included both reactor and solar data of the reference experiments. For LSC at Yemilab (reactor), we have considered a fiducial volume of 2~kton, while for the JUNO reactor, a fiducial volume of 20~kton has been taken into account \cite{JUNO:2020hqc}. In the case of solar neutrinos, we have used a 1~kton  fiducial volume for LSC at Yemilab, and for JUNO, we have considered a fiducial volume of 20~kton for $^7$Be and $hep$ neutrinos, and 16~kton for $^8$B neutrinos (table \ref{table:fiducial}). 
We reconstructed the number of events per bin as a function of energy for reactor detectors of JUNO and Yemilab assuming $\Delta m^2_{21}=7.54 \times 10 ^{-5}$   eV$^2$  (blue) and  $\Delta m^2 _{21}=6.11 \times 10 ^{-5}$   eV$^2$  (red) in Fig.~\ref{noe_reactor}  assuming 10 years of data taking for each experiment. We observed that the oscillation pattern differs between the reference experiments due to the difference in their baselines. This difference holds the potential to enhance the sensitivity in measuring $\Delta m^2_{21}$.

Overall, we have demonstrated the excellent potential of LSC at Yemilab in determining the solar neutrino parameter $\theta_{12}$ from detecting solar neutrinos due to the large number of events that can be attributed to low energy threshold, suppressed number of background, and detection of $pp$ neutrinos (Fig.~\ref{constraints_1d}). 
We have shown the importance of 
detecting $^7$Be solar neutrinos for the JUNO in accurately determining the oscillation parameter $\theta_{12}$. Let us emphasize that the detection of solely $^8$B neutrinos provides limited accuracy (Fig.~\ref{sol11}). However, by combining the $^8$B data with $^7$Be data, the sensitivity can be enhanced to a level comparable to that of LSC at Yemilab. It is important to note that both $^7$Be and $^8$B neutrinos have large flux uncertainties.  Therefore, considering these systematic uncertainties will 
significantly weaken the results.

Combined with the reactor neutrino data coming from the Hanul power plant, LSC at Yemilab we can simultaneously determine the solar neutrino parameters $\theta_{12}$ and $\Delta m^2_{21}$ to a percent level (Fig.~\ref{t12_dm21}).
On the other hand, the reactor neutrino detection in JUNO can precisely determine $\Delta m^2_{21}$, in particular for the best fit value from KamLAND.  Furthermore, for smaller values of $\Delta m^2_{21}$, particularly for the best fit value obtained from solar data
at SK/SNO, LSC at Yemilab exhibits superior capability in determining the precise value of $\Delta m^2_{21}$ compared to JUNO (Fig.~\ref{interval}). 
It is intriguing that the combination of all of those observations, i.e., the solar with reactor data in LSC at Yemilab and the reactor data in JUNO, renders us to simultaneously determine the parameters $\theta_{12}$ and $\Delta m^2_{21}$ {\it with sub-percent level precision} for both cases of the best fit values of KamLAND and SK/SNO. 
We expect such precise determination of the solar neutrino parameters can shed light in probing new physics beyond the SM as well as understanding the neutrino flux and the details of the dynamics of the Sun and the reactors.

\subsection*{Acknowledgments}
Authors are grateful to Stephen Parke and Peter Denton for useful remarks. The work of P.B., M.R. and S.S. is supported by the National Research Foundation of Korea (Grants No. NRF 2020R1I1A3072747 and No. NRF-2022R1A4A5030362). 
S.H.S. is supported by the NRF grant funded by the Korea Ministry of Science and ICT (MSIT) (No. 2017R1A2B4012757, IBS-R016-D1-2019-b01) and IBS-R016-D1.  
S.S. is also supported by IBS-R018-D1.



\bibliography{ref}



\end{document}